\documentclass[a4paper,fleqn,usenatbib]{mnras}

\usepackage{newtxtext,newtxmath}
\usepackage[english]{babel}

\usepackage[english]{babel}

\usepackage[T1]{fontenc}
\usepackage{ae,aecompl}
\usepackage{pdflscape}
\usepackage{subfigure}
\usepackage{hyperref}
\usepackage{graphicx}	
\usepackage{amsmath}	
\usepackage{amssymb}	
\usepackage[flushleft]{threeparttable}

\def \aj {AJ}
\def \apj{ApJ}
\def \apjl {ApJL}

\def \aap {A\&A}
\def \aara {A\&A Rev.}

\def \mnras {MNRAS}

\def \ssr {Space Science Reviews}
\def \spie {Society of Photo-Optical Instrumentation Engineers (SPIE) Conference Series}
\def \apss {Ap\&SS}

\usepackage{amsmath} 
\newcommand{\aox}{$\alpha_{ox}$}
\newcommand{\gam}{$\Gamma \,$}
\newcommand{\angstrom}{\text{\normalfont\AA}}

\newcommand{\sk}{\vskip -8pt}

\title[X-ray properties of z\textgreater4 blazars]{X-ray properties of z\textgreater4 blazars}

\author[Ighina et al.]{L. Ighina$^{1,2}$\thanks{E-mail: l.ighina@campus.unimib.it}, A. Caccianiga$^1$,  A. Moretti$^1$, S. Belladitta$^{1,3}$, R. Della Ceca$^1$, 
\newauthor  L. Ballo$^{4}$, D. Dallacasa$^{5,6}$
\\
$^{1}$INAF - Osservatorio Astronomico di Brera, via Brera 28, I-20121 Milan, Italy\\
$^{2}$Dipartimento di Fisica G. Occhialini, Università di Milano-Bicocca, Piazza della Scienza 3, I-20126 Milano, Italy\\
$^{3}$DiSAT, Università degli Studi dell’Insubria, Via Valleggio 11, 22100, Como, Italy\\
$^{4}$XMM-Newton Science Operations Centre, ESAC/ESA, PO Box 78, E-28691 Villanueva de la Ca$\tilde{n}$ada, Madrid, Spain 0000-0002-5036-3497\\
$^{5}$Dipartimento di Astronomia, Università di Bologna, via Ranzani 1, 40127, Bologna, Italy\\
$^{6}$INAF - Istituto di Radioastronomia, via Gobetti 101, I-40129 Bologna, Italy
}

\date{}

\pubyear{2019}

\begin{document}
\label{firstpage}
\pagerange{\pageref{firstpage}--\pageref{lastpage}}
\maketitle
\sk
\sk
\begin{abstract}
We present the X-ray analysis of the largest flux-limited complete sample of blazar candidates at z\textgreater4 selected from the Cosmic Lens All Sky Survey  (CLASS). After obtaining a nearly complete (24/25) X-ray coverage of the sample (from Swift-XRT, XMM-Newton and Chandra), we analysed the spectra in order to identify the bona-fide blazars. We classified the sources based on the shape of their Spectral Energy Distributions (SEDs) and, in particular, on the flatness of the X-ray emission and its intensity compared to the optical one. We then compared these high-z blazars with a blazar sample selected at lower redshifts ($\bar{z}\sim$1). We found a significant difference in the X-ray-to-radio luminosity ratios,  with the CLASS blazars having a mean ratio 2.4$\pm$0.5 times larger than low-z blazars. We tentatively interpret this evolution as due to the interaction of the electrons of the jet with the Cosmic Microwave Background (CMB) photons, which is expected to boost the observed X-ray emission at high redshifts. Such a dependence has been already observed in highly radio-loud AGNs in the recent literature. This is the first time it is observed using a statistically complete radio flux limited sample of blazars. We have then evaluated whether this effect could explain the differences in the cosmological evolution recently found between radio and X-ray selected samples of blazars. We found that the simple version of this model is not able to solve the tension between the two evolutionary results. \\
\end{abstract}




\section{Introduction}

The detection and the study of very high redshift Active Galactic Nuclei (AGNs) is the best way to provide observational constraints to current theoretical models of SMBHs growth from primordial seed BHs (e.g. \citealt{Volonteri2010}). However, the number of high-z AGNs observed could be largely affected by obscuration effects which are difficult to quantify (e.g.  \citealt{Zeimann2011}). This produces a systematic uncertainty in the statistical estimates due to the assumptions made about the absorbed population. This is why, in the last years, the class of blazars has acquired a particular importance (e.g. \citealt{Volonteri2011} and \citealt{Sbarrato2015}). These objects are radio-loud (RL\footnote{i.e. with a radio-loudness  $R$ \textgreater \, 10, where $R$ = (f$_{5\,GHz}$ / f$_{4400\,\angstrom})$, \cite{Kellerman1989}. }) AGNs whose relativistic jet points directly towards us, making obscuration less important, because we are observing them roughly perpendicularly to the dusty torus. At the same time, from the number of observed blazars it is possible to infer the total density of  RL AGNs with similar properties at a given redshift  (N$_{tot}$ $\approx$ N$_{obs}$ $\times$ 2$\Gamma^2$, e.g. \citealt{Ghisellini2014})\footnote{Where $\Gamma$ is the bulk Lorentz factor of the plasma within the jet. Typically $\Gamma\sim$10-15, e.g. \cite{Ghisellini2010}.}. This estimate is potentially free from the usual bias due to the obscuration and, therefore, it complements the independent estimates based on non-blazar AGNs. However, for a correct application of this method it is fundamental to have a reliable procedure to distinguish blazars from non-blazar sources.\\
To date the largest flux-limited sample of blazar candidates at redshift larger than 4 has been derived from the CLASS survey (26 sources, \citealt{Caccianiga2019}, hereafter C19). From the analysis of their radio spectra, C19 identified the 18 bona-fide blazars (i.e. those with a flat radio spectrum) and then derived, for the first time, the density distribution of  blazars at z\textgreater 4. Nevertheless, the analysis of the radio spectra alone is not 100\% reliable in the recognition of all the blazar sources in a sample. As discussed by C19, for instance, there is at least one striking case of high-z quasar (QSO) (J090631+693027, at z=5.47, \citealt{Romani2004}), that shows a peaked radio spectrum, apparently not supporting a blazar classification. However, several pieces of evidence, based on Very Long Baseline Interferometer (VLBI) data \citep{Coppejans2017}, variability arguments and on the shape of the SED, have clearly revealed the blazar nature of this object (see discussion in C19). This demonstrates that the analysis of the radio spectra can provide a simple tool to quickly classify the sources in large samples, but it can miss a fraction of blazars. An X-ray analysis can provide a more accurate tool to discriminate blazars from non-blazars. Indeed, as described in the following sections, blazars are characterised by a strong, with respect to the optical, and flat X-ray emission which, combined with the other pieces of information, can then be used for a more reliable classification. In this paper we present the analysis of the X-ray observations of the CLASS sample together with  their classification (blazar vs non-blazar).\\
In addition, a systematic study of the X-ray properties of the CLASS high-z blazars can help to understand the discrepancy found between the cosmological evolution of the radio-selected blazars at z>4 and the X-ray selected ones. In particular, in C19, we found that the space density of the blazars with a radio luminosity between $\sim$ 10$^{43}$ and 3x10$^{44}$ $erg$ $s^{-1}$ (at 1.4 GHz) is in good agreement with the predictions recently presented by \cite{Mao2017}, suggesting a peak at redshift $\sim$2. This is significantly different from what has been found by \cite{Ajello2009} using an X-ray selected sample of blazars and that suggests a much higher redshift peak (z$\sim$4). If the average X-ray-to-radio luminosity ratio in blazars is constant along the cosmic time, the observed differences are difficult to explain. On the contrary, a dependence of this luminosity ratio with redshift could, in principle, reconcile the two results. For this reason, in this paper we will use the results of our X-ray analysis to compare the X-ray-to-radio luminosity ratio of the high-z blazars in CLASS with a reference sample selected at lower redshift, searching for any possible dependence.\\
This paper is organised as follows: in section \ref{sec:sample} we summarise the selection criteria of the CLASS sample, while in section \ref{sec:xrayan} we report the X-ray observations of the sources and their analysis. The SED of each source is reported in section \ref{sec:seds}. In section \ref{sec:blaid} we use the results of the analysis to identify the blazar-like objects and in section \ref{sec:xrayradio} we compare their X-ray and radio properties with low-z blazars. Finally, in section \ref{sec:sum} we summarise our results.\\
Through this paper we assume a flat $\Lambda$CDM cosmology with H$_0$=70 km s$^{-1}$ Mpc$^{-1}$, $\Omega_{\Lambda}$=0.7,  $\Omega_{M}$=0.3. Spectral indexes are given assuming S$_{\nu}$ $\propto$ $\nu^{-\alpha}$. All the errors are reported at 90\% confidence level, unless otherwise specified.

\section{The CLASS sample of high redshift blazars}
\label{sec:sample}
\begin{figure}
\centering
\includegraphics[width=\columnwidth]{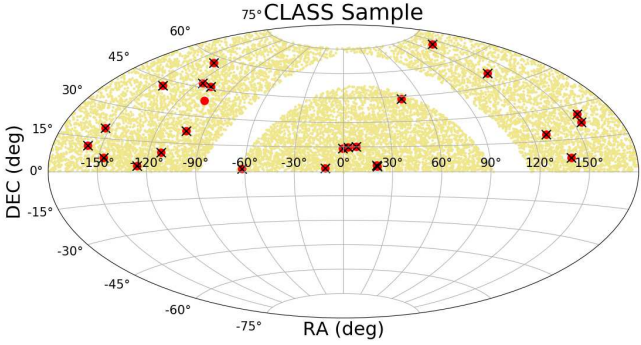}
\caption{ Sky coverage of the CLASS survey (yellow points). We report in red the z$\textgreater$4 sources confirmed by C19 and with the black crosses those sources for which  X-ray data are available.}
\label{sky}
\end{figure}

All the sources analysed in this work have been selected  from the CLASS survey (\citealt{myers2003,Browne2003}), a density flux limited survey (S$_{5GHz}$ \textgreater 30 mJy) of flat spectrum radio sources, which covers most of the northern sky (16300 deg$^{2})$ and that contains more than 11000 objects (see Fig \ref{sky}). It was built by combining the NRAO VLA Sky Survey (NVSS) at 1.4 GHz \citep{condon1998} with the Green-Bank Survey (GB6) at 5 GHz (\citealt{gregory1996}) and by selecting only those objects with a flat (-0.5 \textless$ \alpha $ \textless0.5) spectrum between 1.4 and 5 GHz.\\
The selection of the high-z sources in the CLASS survey has been described in details in C19. Here we summarise the main steps. The optical counterparts of the CLASS sample have been searched using the Panoramic Survey Telescope and Rapid Response System (Pan-STARRS1, PS1, \citealt{Chambers2016}) an optical survey in five different filters ($g$, $r$, $i$, $z$, $Y$), within a search radius of 0.6 arcsec from the radio position. Then, we have used the PS1 photometric data and the \textit{dropout} method to efficiently pre-select objects at high redshift (4\textless z\textless6). In order to confirm the high-z nature of these selected objects we have carried out a systematic spectroscopic follow-up. In particular, all the candidates without an archival optical spectrum were observed at the Large Binocular Telescope (LBT) or  at the Telescopio Nazionale Galileo (TNG), providing a spectroscopic estimate of the redshift for all the sources.\\ 
Most of the observed objects were confirmed as high-z AGN leading to a final sample of 26 confirmed z\textgreater4 AGNs. However, after the publication of C19, we discovered that one of the sources (GB6J160608+312504) has an incorrect spectroscopic redshift reported in the literature (see \citealt{Belladitta2019} for further information). For this reason, we do not consider that object in this work, reducing the sample to 25 sources (see red dots in Fig \ref{sky}). In C19 we extended the range of radio frequencies used to define the CLASS survey to refine the spectral classification of each high-z object, keeping as blazars only the 18 sources with a flat spectrum between 150 MHz up to 8.4 GHz (observed frame). The remaining objects show a possibly peaked spectrum that does not support their blazar nature. However, as already pointed out, some blazars can show a peaked spectrum. This is in part due to the non simultaneous radio data used for the classification, that may lead to mis-classify a variable blazar as non-flat source. In addition, there is at least one known case of a blazar (the already mentioned J090631+693027) that shows a peaked radio spectrum even using simultaneous radio data. It is also known that flaring blazars may temporarily show a peaked spectrum, like the High Frequency Peakers \citep{Orienti2007,Orienti2010}. The X-ray analysis discussed in the next sections will help us to derive a firmer classification of all the 25 high-z blazar candidates.

\section{X-ray Spectral Analysis}
\label{sec:xrayan} 

\begin{figure}
\centering
\includegraphics[width=\columnwidth]{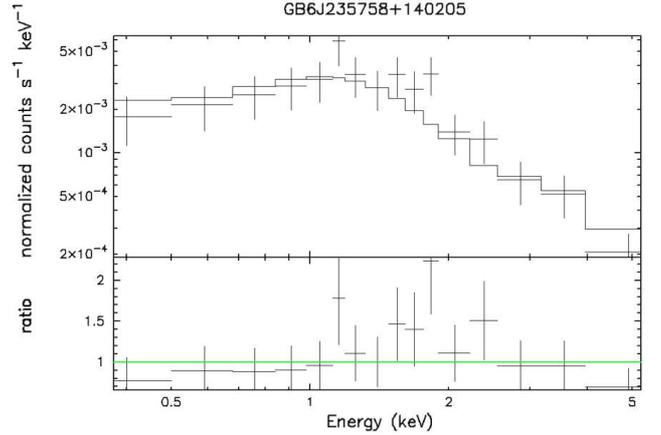}
\caption{Top panel: Swift-XRT  X-ray spectrum of GB6J235758+140205, example of a good blazar candidate. The model used is a Galactic absorbed power law with parameters: N$_H$ = 3.41x10$^{20}cm^{-2}$ and \gam = 1.50 $\pm$ 0.18. Bottom panel: ratio between the data and the model.}
\label{spec}
\end{figure}

\begin{figure}
\centering
\includegraphics[width=\columnwidth]{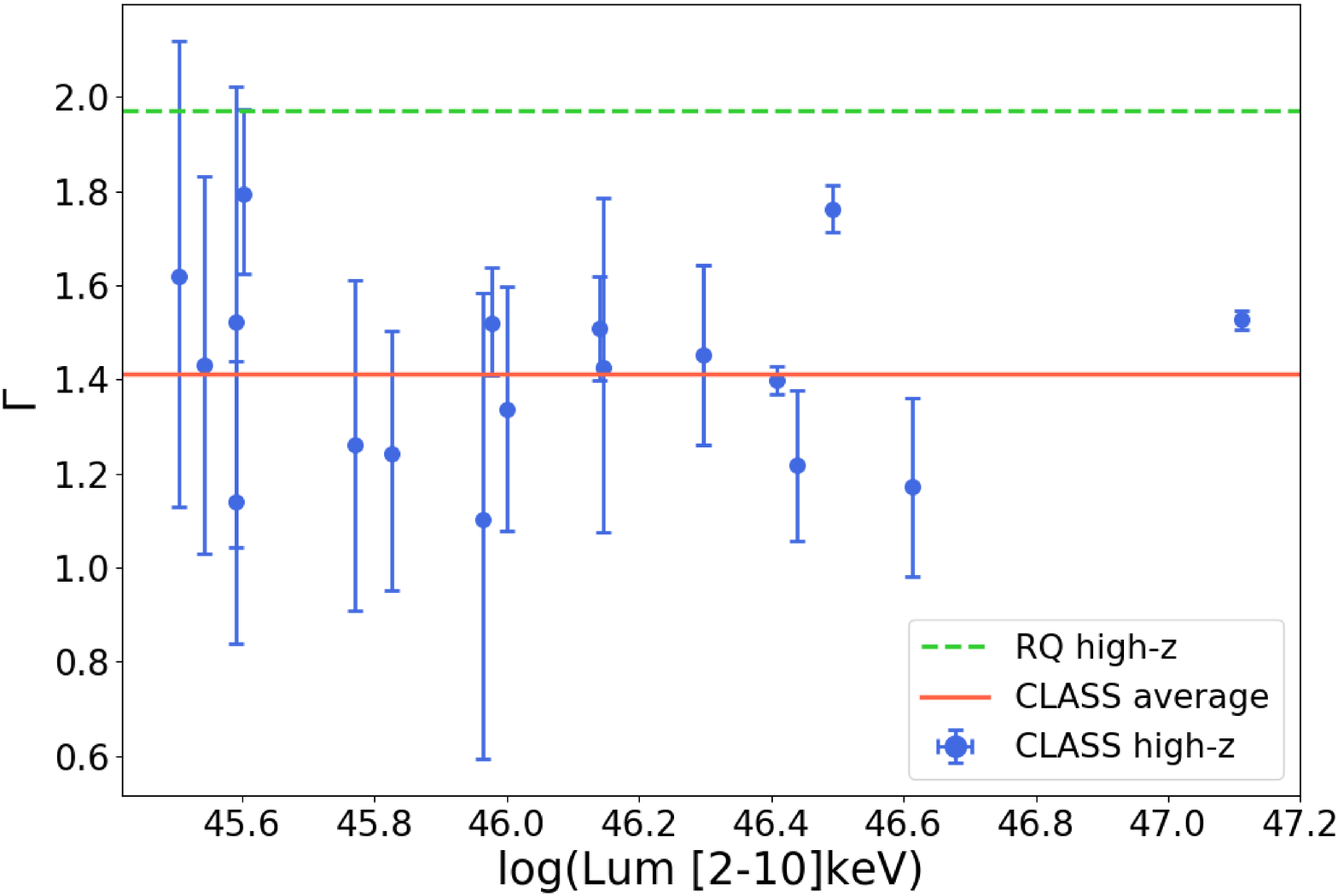}
\caption{Photon index versus the rest frame luminosity [2-10] keV for the sources of the CLASS sample with a reasonable estimate on the photon index (error\textless 0.5). The red line represents the average value of the CLASS sample, $\bar{\Gamma}=1.41$, whereas the green dashed line is the average value of the sample of RQ AGNs discussed in Shemmer et al. (2005), $\bar{\Gamma}=1.97$. }
\label{gavslum}
\end{figure}

Out of the 25 high-z sources in the CLASS sample, 16 have X-ray data available in the public archives of  XMM-\textit{Newton}, Chandra and Swift-XRT. In order to complete the X-ray coverage of the sample, we carried out a dedicated Swift-XRT follow-up of the remaining 9 objects. Only one of them (GB6J171103+383016) has not been observed yet and, therefore, it will not be included in this work. To date we have X-ray data for 24 high-z objects (see Table \ref{Tab1} and black crosses in Fig \ref{sky}).\\
\begin{table*}
\centering
\footnotesize
\tabcolsep 4pt
\caption{X-ray observations of the CLASS sample.}
\begin{tabular*}{\textwidth}{lcccccrc}
\hline
\hline
\textbf{Name} &\textbf{z}&\textbf{Coordinates} &\textbf{X-ray}&\textbf{Observation} &\textbf{Observation}  &\textbf{Exp. time}&\textbf{Ref.}  \\
&&\textbf{(J2000.0)}&\textbf{Telescope}&\textbf{ID}& \textbf{Date}&\textbf{(ks)}\\
\hline
\hline
\\
GB6J001115+144608	&4.96	&00 11 15.24 +14 46 01.8	&Chandra	&3957			&2003/05					&3.49			&3		\\
&&														&XMM		&0600090101	&2010/01					&29.20			&8		\\	
GB6J003126+150729	&4.29	&00 31 26.80 +15 07 39.5	&Chandra	&18442			&2016/06					&5.35			&2		\\
GB6J012126+034646	&4.13	&01 21 26.15 +03 47 06.7	&Chandra	&3151 	    		&2002/02					&5.68			&1,7	\\
&&														&XMM		&0200730301	&2004/01					&21.60			&4		\\	
GB6J012202+030951	&4.00	&01 22 01.91 +03 10 02.4	&Swift	   	&00035924	   	&2007/06 - 2008/01 			&4.03					\\
&&														&Swift		&00036780		&2008/01					&4.65					\\	
GB6J025758+433837	&4.07	&02 57 59.08 +43 38 37.7	&Chandra	&18449	    		&2015/12					&5.51			&2		\\
GB6J083548+182519	&4.41	&08 35 49.43 +18 25 20.1	&Swift		&00087221		&2017/01-02-05				&49.50  			&2  		\\
GB6J083945+511206	&4.40	&08 39 46.22 +51 12 02.8	&Chandra	&3562			&2004/01					&4.9			&5,8,9	\\
&&														&Swift		&00515375		&2012/02					&29.75 			&5		\\	
&&														&XMM		&0650340201	&2011/04					&9.90					\\	
&&														&XMM		&0301340101	&2006/04					&4.90			&8		\\	
GB6J090631+693027	&5.47	&09 06 30.75 +69 30 30.8	&Chandra	&5637			&2005/07					&29.79			& 1,10	\\
&&														&Swift		&00035974		&2006/10 - 2014/02			&30.46					\\	
&&														&Swift		&00035369		&2006/01-05-06				&5.90					\\	
GB6J091825+063722	&4.22	&09 18 24.38 +06 36 53.4	&Chandra	&3563	    		&2002/12					&4.90			&5,9	\\
GB6J102107+220904	&4.26	&10 21 07.58 +22 09 21.6	&XMM		&0406540401	&2016/11					&22.70  			&2 		\\
GB6J102623+254255	&5.28	&10 26 23.62 +25 42 59.4	&Chandra	&12167  		&2012/03					&4.99  			&1		\\
&&														&Swift		&000325000		&2012/06					&10.11			&6		\\	
GB6J132512+112338	&4.42	&13 25 12.49 +11 23 29.8	&Chandra	&3565	    		&2003/03					&4.70			&1,9	\\
GB6J134811+193520	&4.40	&13 48 11.26 +19 35 23.5	&Swift		&00087542		&2017/11-12 - 2018/01		&46.79  			&2 		\\
GB6J141212+062408	&4.47	&14 12 09.97 +06 24 06.8	&Chandra	&12169    		&2011/03					&4.1			&1		\\
&&														&Swift		&00085421		&2014/04-05-06-08-12		&26.38					\\	
GB6J143023+420450	&4.72	&14 30 23.74 +42 04 36.5	&Chandra  	&7874		  	&2007/03  					&10.57			&11		\\
&&														&Swift		&00080752		&2014/07					&9.61					\\	
&&														&Swift		&00036798		&2013/11-12					&4.27					\\	
&&														&XMM		&0212480701	&2005/06					&11.00			&1,12	\\	
&&														&XMM		&0111260101	&2002/12					&9.90			&13,14	\\	
&&														&XMM		&0111260701	&2003/01					&11.50			&14,15	\\	
GB6J151002+570256	&4.31	&15 10 02.93 +57 02 43.4	&Chandra	&2241    		&2001/06					&88.98			&1,16,17\\
&&														&XMM		&0111260201	&2002/05					&8.70			&13,15,18\\	
GB6J153533+025419	&4.39	&15 35 33.88 +02 54 23.4	&Swift	    	&00087222	   	 &2016/12 - 2017/01  	 	&26.40			&2		\\
GB6J161216+470311	&4.36	&16 12 16.76 +47 02 53.6	&Swift		&00087543		&2017/11-12					&48.68   		&2		\\
GB6J162956+095959 &5.00	&16 29 57.28 +10 00 23.5	& Swift		&03109568		&2019/02-03-04-05-06		&18.10			&		\\
GB6J164856+460341	&5.36	&16 48 54.53 +46 03 27.4	&Swift		&00010651		&2018/04					&14.40  					\\
GB6J171521+214547	&4.01	&17 15 21.25 +21 45 31.7	&Chandra	&4815	    		&2004/06					&9.54			&1,7	\\
GB6J195135+013442	&4.11	&19 51 36.02 +01 34 42.7	&Swift		&00036263		&2007/03					&10.23 			&1,19 	\\
GB6J231449+020146	&4.11	&23 14 48.71 +02 01 51.1	&Chandra	&18448	    		&2016/01					&5.88			&2		\\
&&														&Swift		&00085422		&2014/11 - 2015/01-04-05	&24.90					\\	
GB6J235758+140205	&4.35	&23 57 58.56 +14 02 01.9	&Swift		&00087544		&2017/11-12 - 2018/01		&34.10   				\\
\\

\hline
\hline
\end{tabular*} 
\label{Tab1}
\begin{tablenotes}
\item All the sources composing the CLASS sample for which X-ray observations are available, with their redshift and sky coordinates. For each observation we report the satellite, the identification number, the observation date and the exposure time. The last column reports other works where the X-ray observations have also been analysed: 1 = \cite{Wu2013}, 2 = \cite{Zhu2018}, 3 = \cite{Shemmer2006}, 4 = \cite{Shemmer2005}, 5 = \cite{Sbarrato2013}, 6 = \cite{Sbarrato2012}, 7 = \cite{Vignali2003}, 8 = \cite{Saez2011} 9 = \cite{Bassett2004}, 10 = \cite{Romani2004}, 11 = \cite{Cheung2012} , 12 = \cite{Eitan2013}, 13 = \cite{Grupe2005}, 14 = \cite{Worsley2004}, 15 = \cite{Page2005}, 16 = \cite{Siem2003}, 17 =  \cite{Yuan2003}, 18 = \cite{Yuan2006}, 19  = \cite{Healey2008}.
\end{tablenotes}

\end{table*}
In order to have a systematic and up-to-date analysis of the entire sample, we carried out the X-ray analysis of all the sources, both the newly observed and the ones already discussed in the literature.\\
Data of the Swift-XRT telescope \citep{burrows2005} were downloaded from HEASARC public archive. They were then processed with the specific Swift software included in the package HEASoft (v. 6.23, \citealt{evans2009}). Chandra observations, made through the  Advanced CCD Imaging Spectrometer (ACIS; \citealt{Garmire2003}), were reduced using the software CIAO (v4.6\footnote{We checked that the results obtained with the latest version (v4.11) do not differ from ours.}). Source and background counts were extracted using \texttt{SPECEXTRACT}, where the two regions consisted in circles of radii $\sim$2"  and $\sim$30" respectively, with all the background regions, chosen close to the target, not containing other X-ray sources. For XMM observations we only considered the data from the PN detector\footnote{We did not consider the data from the MOS detectors since the EPIC-pn observations alone provided enough photon counts for the purposes of our analysis.} (\citealt{struder2001}), which were reduced through standard XMM-Newton Science Analysis System (SAS; v.17.0) routine. The photon counts this time have been extracted from a circle of radius  $\sim$40" for the targets and from a nearby circle of radius $\sim$100" source-free. For all the observations the source regions have been centred in the peak of the X-ray emission.\\
We performed the X-ray spectral analysis for each observation  independently using the package \texttt{XSPEC} (v.12.9.0i) and by fitting the observed spectra with a simple power law absorbed by the Galactic column density along the line-of-sight. As an example, we report in Fig \ref{spec} the X-ray spectrum of the source GB6J235758+140205, a good blazar candidate. We do not consider intrinsic absorption at the source redshift since, usually, blazars do not show evidence of a significant absorption (e.g. \citealt{Giommi2019}, but see \citealt{Eitan2013} and \citealt{Saez2011} for the discussion of some exceptions). In any case, the available statistics is in general too limited to attempt a search of any extra absorption in our targets and, for the few sources with enough counts, the addition of an extra component in the fitting model does not affect the analysis.\\
For the objects with more than 40 net counts, after grouping them in bins of energy with at least 10 counts each, we used the $\chi^2$ statistics to derive the best-fit, whereas for those sources with a limited number of photon counts ($\lesssim$40, 11 observations), where an efficient grouping was not possible, we performed the fit using the C-statistic \citep{Cash1979} on the data.
In Table \ref{table:analysis} we report the results of the fit, together with the 90\% confidence range for the photon indexes and the observed fluxes. The flux error has been computed considering both the uncertainty on the photon index\footnote{We constrained the value of the photon index within the range 1\textless$\Gamma$\textless2.5, since these are the limit values observed in  flat radio spectrum quasars (FRSQs), e.g. \citealt{Giommi2019}.} and the Poissonian error associated to the photon counts.\\ 
For the objects already published  we found a good agreement, in general, between our results and those found in the literature (see Table \ref{table:analysis}). For about 40\% of the sample, we analysed multiple observations and, even though blazars are known to be highly variable in the X-rays (e.g. \citealt{Giommi2019}), we decided to combine their best-fit values with a weighted average, because there are no striking evidences of variability in our data. The only source that presents a significant variability between different observations is GB6J143023+420450, which has already been deeply studied in the literature (e.g. \citealt{Worsley2004} and \citealt{Page2005}).  For simplicity, however, in the following paragraphs we use the analysis of the combined spectrum also for this object, since we verified that even in this case the main results discussed here would not change.\\
In Fig \ref{gavslum} we report the photon index versus the X-ray luminosity ([2-10] keV rest frame) obtained from the analysis for the majority ($\sim$80\%) of the sources in the sample, i.e. those with a photon index uncertainty \textless0.5. Our sample presents an overall average photon index ($\bar{\Gamma}$=1.41$^{+0.30}_{-0.28}$, red line) significantly flatter than the average value found in \cite{Shemmer2005} for a sample of radio-quiet (RQ) AGNs at z\textgreater4 (green dashed line), supporting the idea that the majority of the sources selected in C19 are indeed blazars.

\begin{table*}

\centering
\footnotesize
\tabcolsep 7pt
\caption{Results of the X-ray analysis}
\begin{center}
\begin{tabular*}{0.8\textwidth}{lcccccccc}

\hline
\hline
\textbf{Name}&\textbf{Obsid(Telescope)}&\boldmath {\(\Gamma\)}&\boldmath{$F_x $}&\boldmath{$log(L_x$)}&\boldmath{\(\chi^2\)} / \textbf{d.o.f.}\\
&&&\textbf{[0.5-10]keV}&\textbf{[2-10] keV}&&\\
\hline
\hline
&&&&&&&\\
GB6J001115+144608	&3957 (C)		 	& 1.71$_{-0.29}^{+0.29}$  	&31.2   	$_{-7.6}^{+9.1}$  	&46.48   &7.9 / 6			\\
					&0600090101 (X)		& 1.76$_{-0.05}^{+0.05}$	&30.0	$_{-1.6}^{+1.7}$	&46.49	&71.6 / 75		\\
GB6J003126+150729	&18442 (C)		 	& 2.50$_{-0.59}^{}$			&2.78	$_{-1.4}^{+4.0}$	&45.61	&19.1 /12*		\\
GB6J012126+034646	&3151 (C)		 	& 1.85$_{-0.33}^{+0.34}$	&8.8 	$_{-1.5}^{+3.0}$	&45.83	&2.2 / 3	 		\\
					&0200730301 (X)		& 1.76$_{-0.27}^{+0.30}$	&4.2 	$_{-0.5}^{+1.2}$	&45.50	&10.8 / 7		\\
GB6J012202+030951	&00035924 (S)	 	& 1.13$_{-0.29}^{+0.28}$	&133.1	$_{-34.6}^{+21.0}$	&46.61	&0.8 / 4			\\
					&00036780 (S)	 	& 1.20$_{-0.25}^{+0.25}$	&149.2	$_{-34.2}^{+28.1}$	&46.70	&3.7 / 7			\\
GB6J025758+433837	&18449 (C)		 	& 1.43$_{-0.35}^{+0.36}$	&30.3	$_{-6.9}^{+7.7}$	&46.15	&1.8 / 6			\\
GB6J083548+182519	&00087221 (S)	 	& 1.34$_{-0.21}^{+0.21}$	&21.0	$_{-4.1}^{+4.3}$	&46.00	&3.9 / 6			\\
GB6J083945+511206	&3562 (C)		 	& 1.65$_{-0.29}^{+0.31}$	&14.4	$_{-4.1}^{+3.8}$	&46.00	&1.6 / 4			\\
					&00515375 (S)	 	& 1.34$_{-0.46}^{+0.46}$	&17.4	$_{-6.8}^{+6.7}$	&45.92	&3.8 / 3			\\
					&0650340201 (X)		& 1.56$_{-0.16}^{+0.18}$	&15.3	$_{-2.0}^{+1.9}$	&45.98	&10.6 / 11		\\
					&0301340101 (X)		& 1.44$_{-0.20}^{+0.21}$	&17.4	$_{-3.1}^{+2.8}$	&45.98	&16.9 / 20		\\
GB6J090631+693027	&5637 (C)		 	& 1.51$_{-0.11}^{+0.12}$	&15.7	$_{-1.6}^{+1.6}$	&46.15	&21.1 / 28		\\
					&00035974 (S) 	 	& 1.84$_{-0.34}^{+0.38}$	&9.4 	$_{-1.5}^{+2.3}$	&46.12	&5.2 / 4			\\
					&00035369 (S)	 	& 1.01$_{-0.70}^{+0.62}$	&23.4	$_{-11.1}^{+6.0}$	&46.00	&11.9 / 14*		\\
GB6J091825+063722	&3563 (C)		 	& 1.26$_{-0.35}^{+0.35}$	&14.7	$_{-5.0}^{+4.4}$	&45.77	&1.6 / 4			\\      
GB6J102107+220904	&0406540401 (X)	 	& 2.26$_{-1.64}^{+0.24}$	&7.9 	$_{-1.3}^{+8.0}$	&46.00	&5.5 / 6		 	\\
GB6J102623+254255	&12167 (C)		 	& 1.29$_{-0.34}^{+0.34}$	&13.0	$_{-3.9}^{+3.3}$	&45.91	&7.0 / 8			\\
					&00032500 (S)	 	& 1.00$_{}^{+0.38}$ 		&14.9 	$_{-5.4}^{+3.5}$	&45.80	&19.8 / 16*		\\
GB6J132512+112338	&3565 (C)	     		& 1.52$_{-0.50}^{+0.51}$	&6.5	$_{-2.5}^{+4.4}$	&45.59 	&0.8 / 2*		\\
GB6J134811+193520	&00087542 (S)	 	& 1.83$_{-0.55}^{+0.49}$	&4.2 	$_{-0.8}^{+2.1}$	&45.55 	&3.7 / 2			\\
GB6J141212+062408	&12169 (C)	     		& 1.57$_{-0.71}^{+0.70}$	&6.4	$_{-2.2}^{+4.9}$	&45.63	&2.9 / 2*		\\
					&00085421 (S)	 	& 1.59$_{-0.90}^{+0.80}$	&4.4 	$_{-2.1}^{+2.7}$	&45.47	&4.99 / 4*		\\
GB6J143023+420450	&7874 (C)  		 	& 1.33$_{-0.05}^{+0.05}$	&235.1	$_{-11.5}^{+11.6}$	&47.10	&87.1 / 63		\\
					&00080752 (S)	 	& 1.14$_{-0.14}^{+0.14}$	&218.1	$_{-38.5}^{+36.0}$	&46.95	&7.2 / 13		\\
					&00036798 (S)	 	& 1.21 $_{-0.18}^{+0.18}$	&219.6	$_{-29.0}^{+29.4}$	&46.99	&10.22 / 6		\\
					&0212480701 (X)		& 1.49$_{-0.04}^{+0.04}$	&180.0	$_{-5.5}^{+5.4}$	&47.08	&125.9 / 142		\\
					&0111260101 (X)		& 1.75$_{-0.14}^{+0.14}$	&202.7	$_{-20.7}^{+20.6}$	&47.27	&11.3 / 16		\\
					&0111260701 (X)		& 1.67$_{-0.04}^{+0.04}$	&169.0	$_{-4.8}^{+4.7}$	&47.15	&150.9 / 151		\\
GB6J151002+570256	&2241 (C)		 	& 1.40$_{-0.03}^{+0.03}$	&53.5	$_{-1.7}^{+1.7}$	&46.40	&387.6 / 200		\\
					&0111260201 (X)  	& 1.57$_{-0.09}^{+0.10}$	&47.5	$_{-3.6}^{+3.5}$	&46.47	&121.2 / 102		\\
GB6J153533+025419	&00087222 (S)	 	& 1.22$_{-0.16}^{+0.16}$	&64.0	$_{-9.1}^{+8.8}$	&46.44	&10.4 / 14		\\
GB6J161216+470311	&00087543 (S)	 	& 1.89$_{-0.66}^{+0.70}$	&1.6 	$_{-0.7}^{+1.1}$	&45.15	&1.36 / 4*		\\
GB6J162956+095959 &03109568(S)		& 1.69$_{-0.67}^{+0.62}$	&8.2	$_{-3.1}^{+6.1}$	&45.90	&3.06 / 4*		\\
GB6J164856+460341	&00010651 (S)	 	& 1.09$_{-0.09}^{+1.69}$	&3.1 	$_{-2.6}^{+1.5}$	&45.16	&5.6 / 5	*		\\
GB6J171521+214547	&4815 (C)		 	& 1.14$_{-0.30}^{+0.30}$	&12.4	$_{-3.6}^{+2.0}$	&45.59	&6.7 / 6			\\
GB6J195135+013442	&00036263 (S)	 	& 1.10$_{-0.50}^{+0.49}$	&29.8	$_{-9.6}^{+5.6}$	&45.96	&1.4 / 3*		\\
GB6J231449+020146	&18448 (C)		 	& 1.28$_{-0.51}^{+0.49}$	&12.4	$_{-4.7}^{+2.9}$	&45.68	&3.2 / 3			\\
					&00085422 (S)	 	& 1.86$_{-0.65}^{+0.67}$	&3.5 	$_{-1.3}^{+1.9}$	&45.42	&24.9 / 21*		\\
GB6J235758+140205	&00087544 (S)	 	& 1.50$_{-0.18}^{+0.18}$	&34.5	$_{-5.6}^{+5.1}$	&46.30	&11.5 / 14		\\
\\
\hline                 
\hline

\end{tabular*}
\end{center}
\begin{tablenotes}
	\item \textbf{column 1:} Object name; \textbf{column 2:} Obsid and the telescope (S = Swift, C = Chandra, X = XMM);  \textbf{columns 3:} Photon index with its error. The best fit has been constrained to the interval [1-2.5]; \textbf{column 4:}  X-ray observed flux in the energy range  [0.5-10] keV, in units of $10^{-14}$ erg $cm^{-2}s^{-1}$, with its error; \textbf{column 5:} Logarithm of the rest frame intrinsic Luminosity [2-10] keV in units of $erg$ $s^{-1}$;  \textbf{column 6:} Value of the $\chi^2$ with the respective degrees of freedom. The "*" sign indicates the observations where we adopted the C-statistic. In these cases the values reported correspond to the C-parameter \citep{Cash1979} and the degrees of freedom.
\end{tablenotes}
\label{table:analysis}                                                                                 
\end{table*} 

\section{Spectral energy distributions}
\label{sec:seds}

\begin{figure*}
\centering
\subfigure{\label{1535}\includegraphics[width=0.48\linewidth]{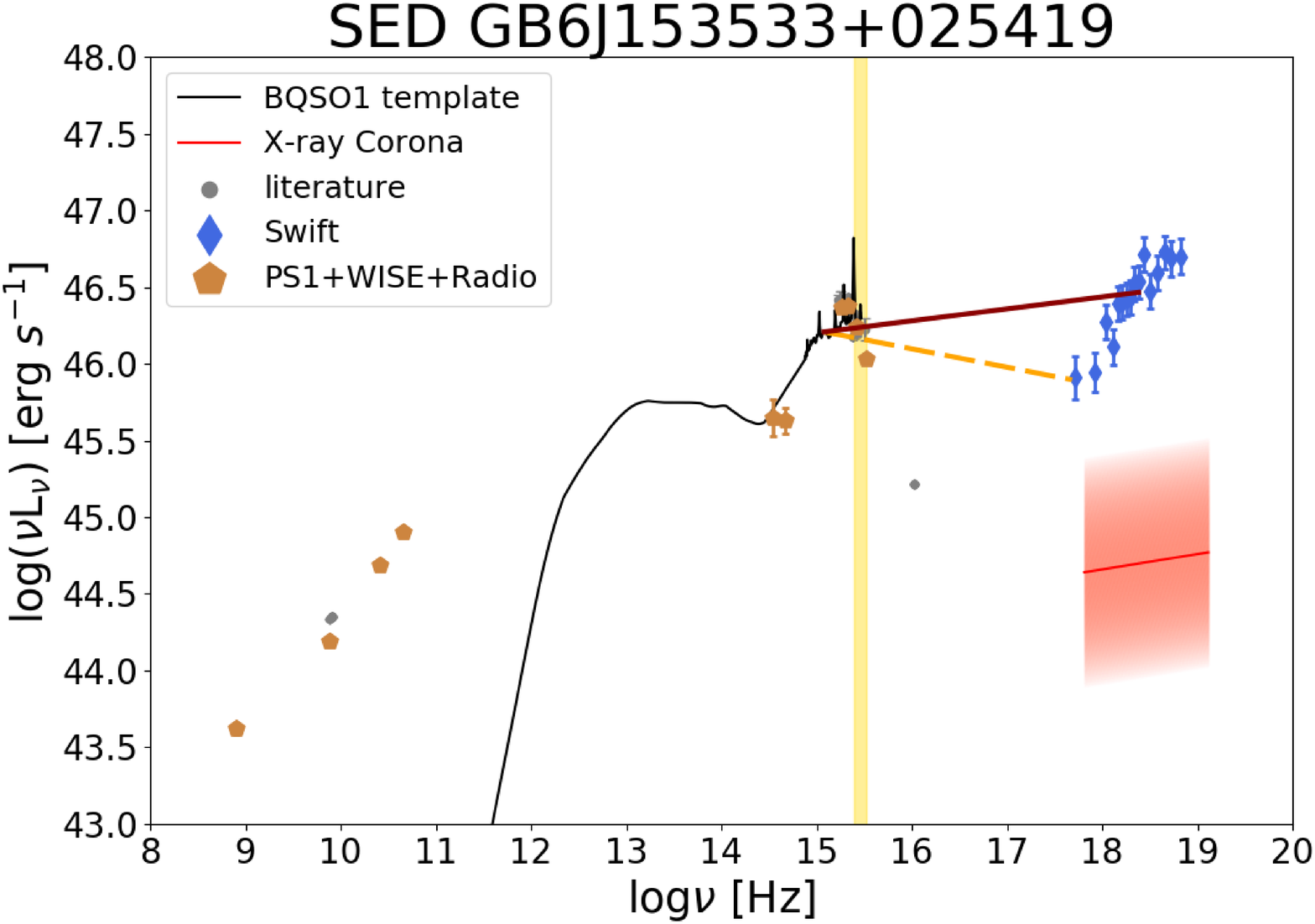}}
\subfigure{\label{1612}\includegraphics[width=0.48\linewidth]{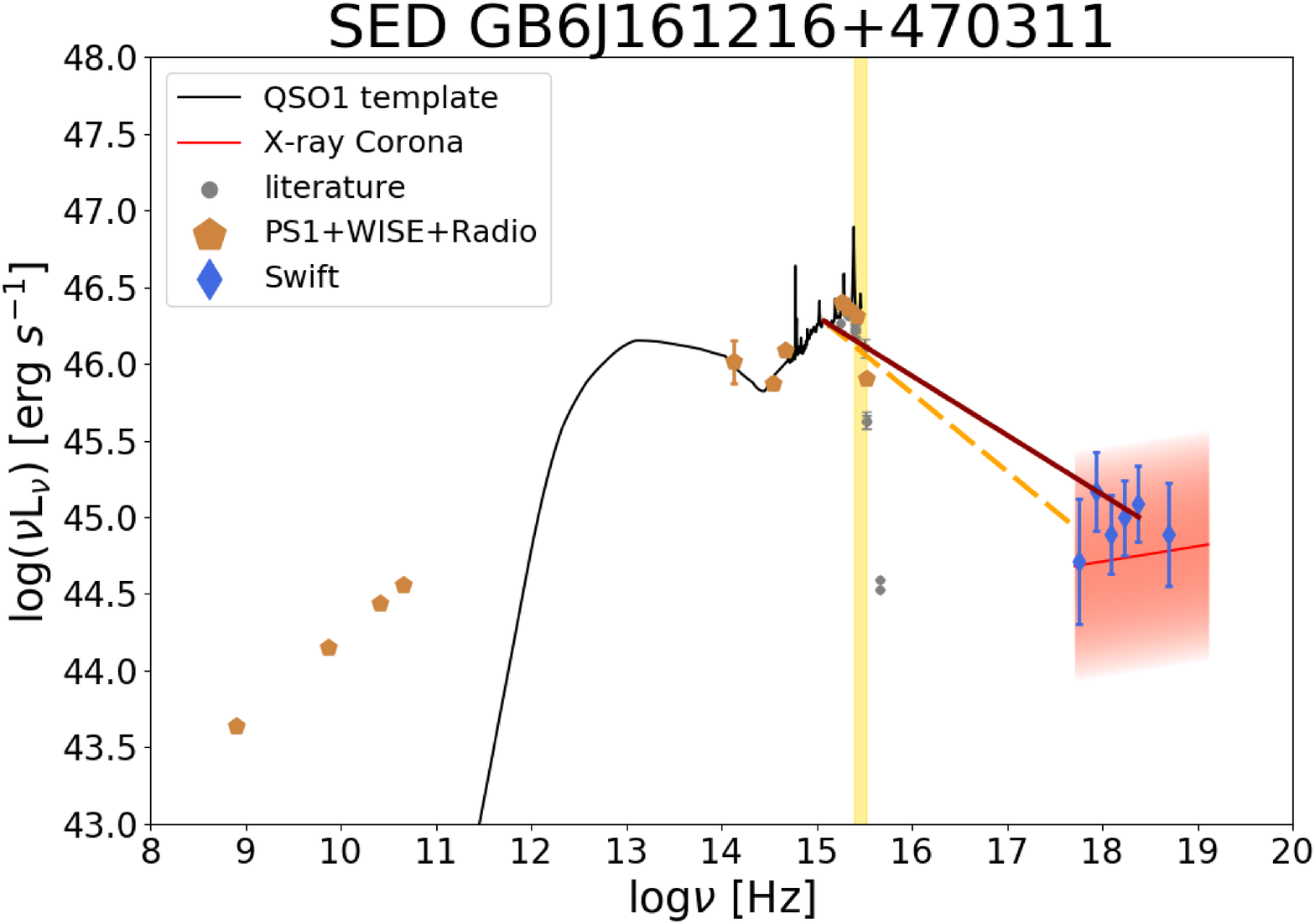}}
\sk
\subfigure{\label{1629}\includegraphics[width=0.48\linewidth]{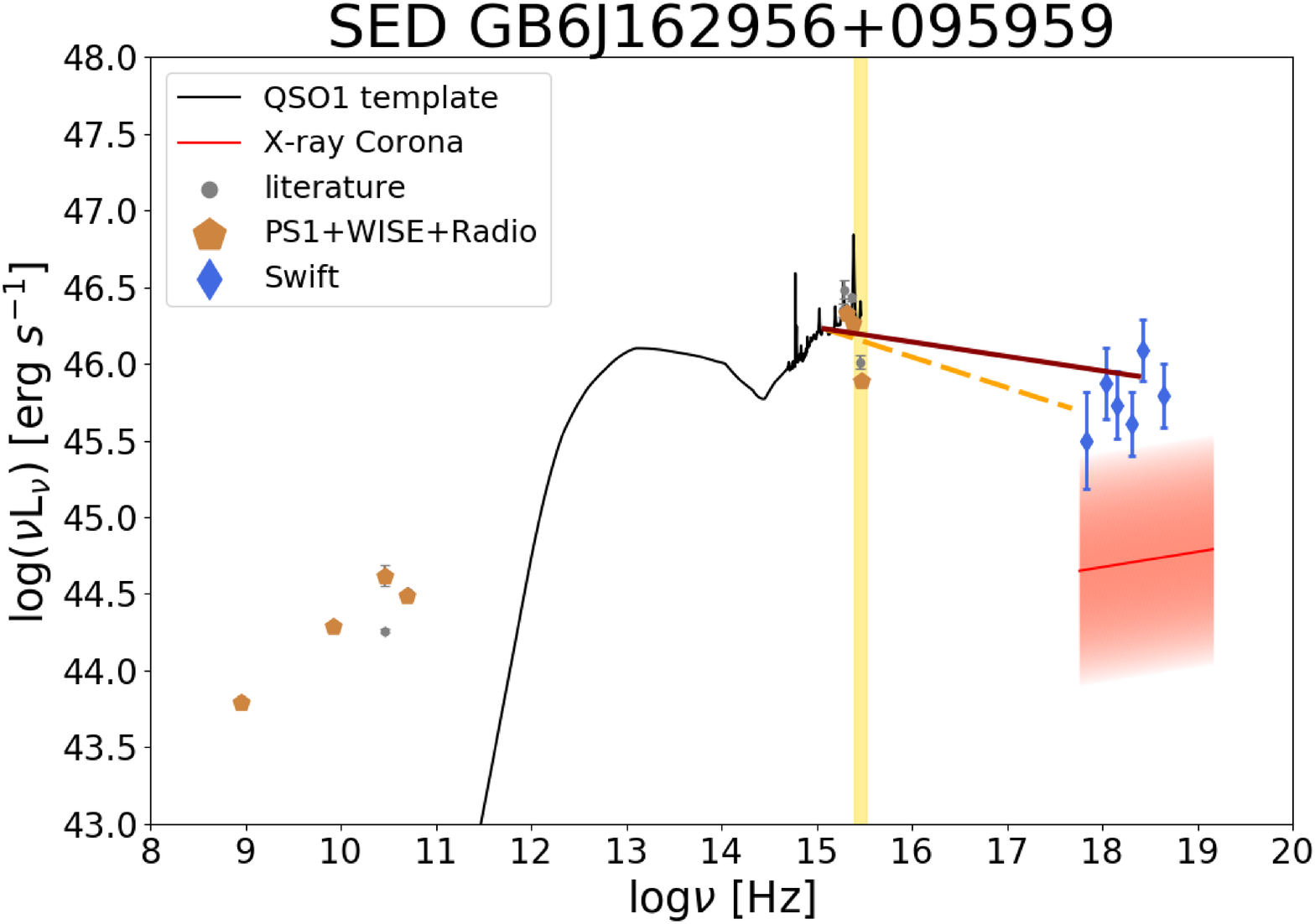}}
\subfigure{\label{1648}\includegraphics[width=0.48\linewidth]{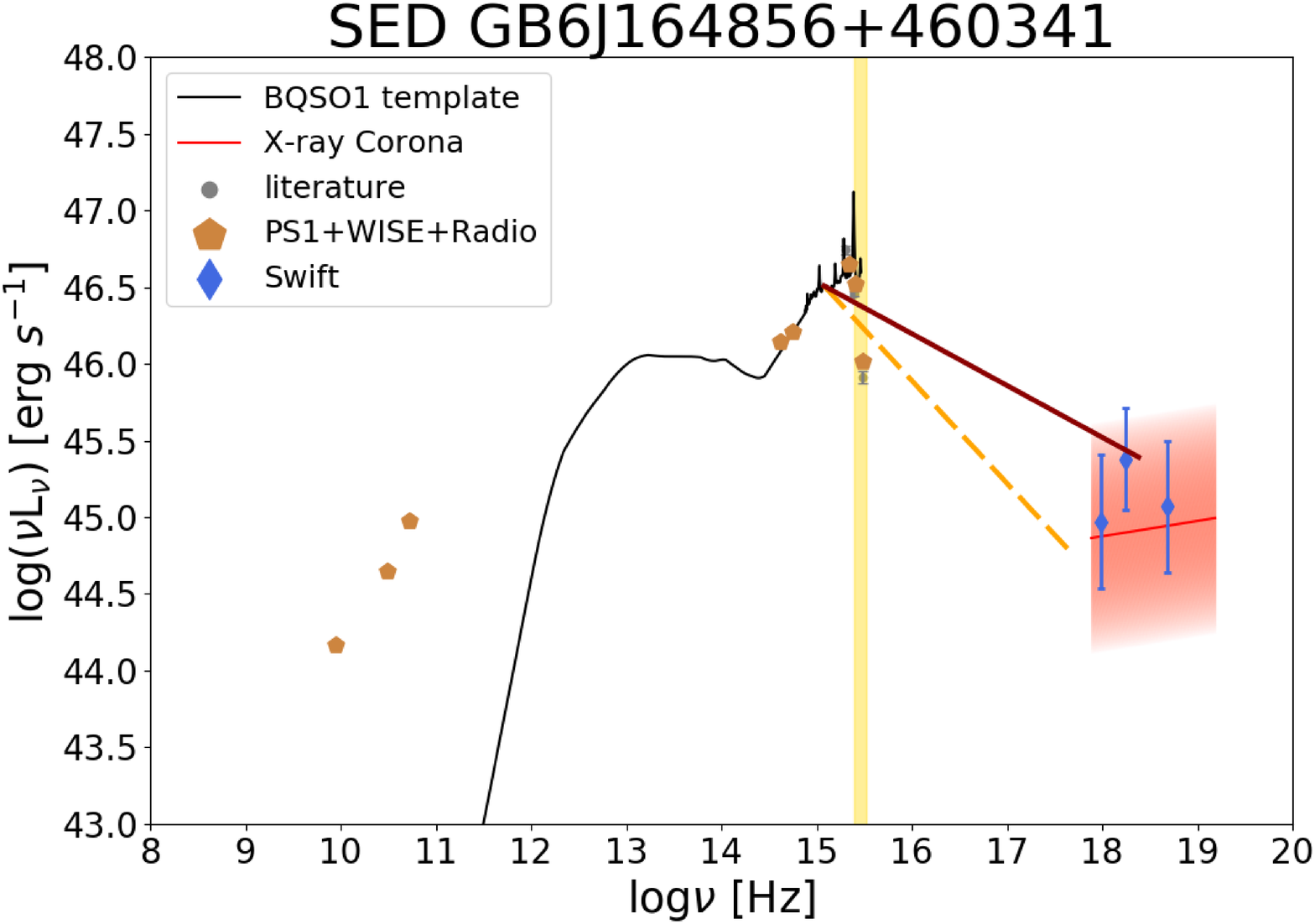}}
\sk
\subfigure{\label{1951}\includegraphics[width=0.48\linewidth]{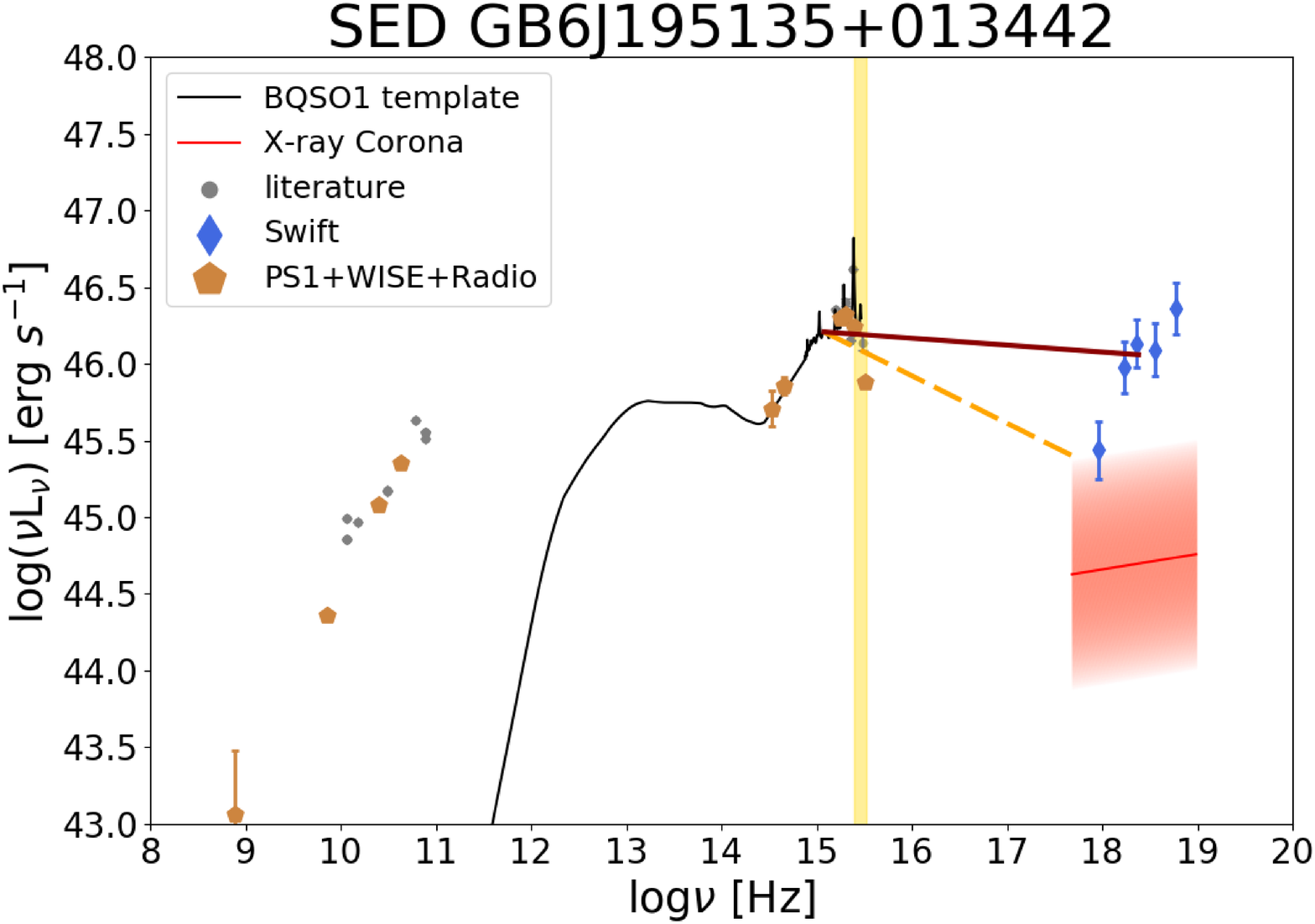}}
\subfigure{\label{1715}\includegraphics[width=0.48\linewidth]{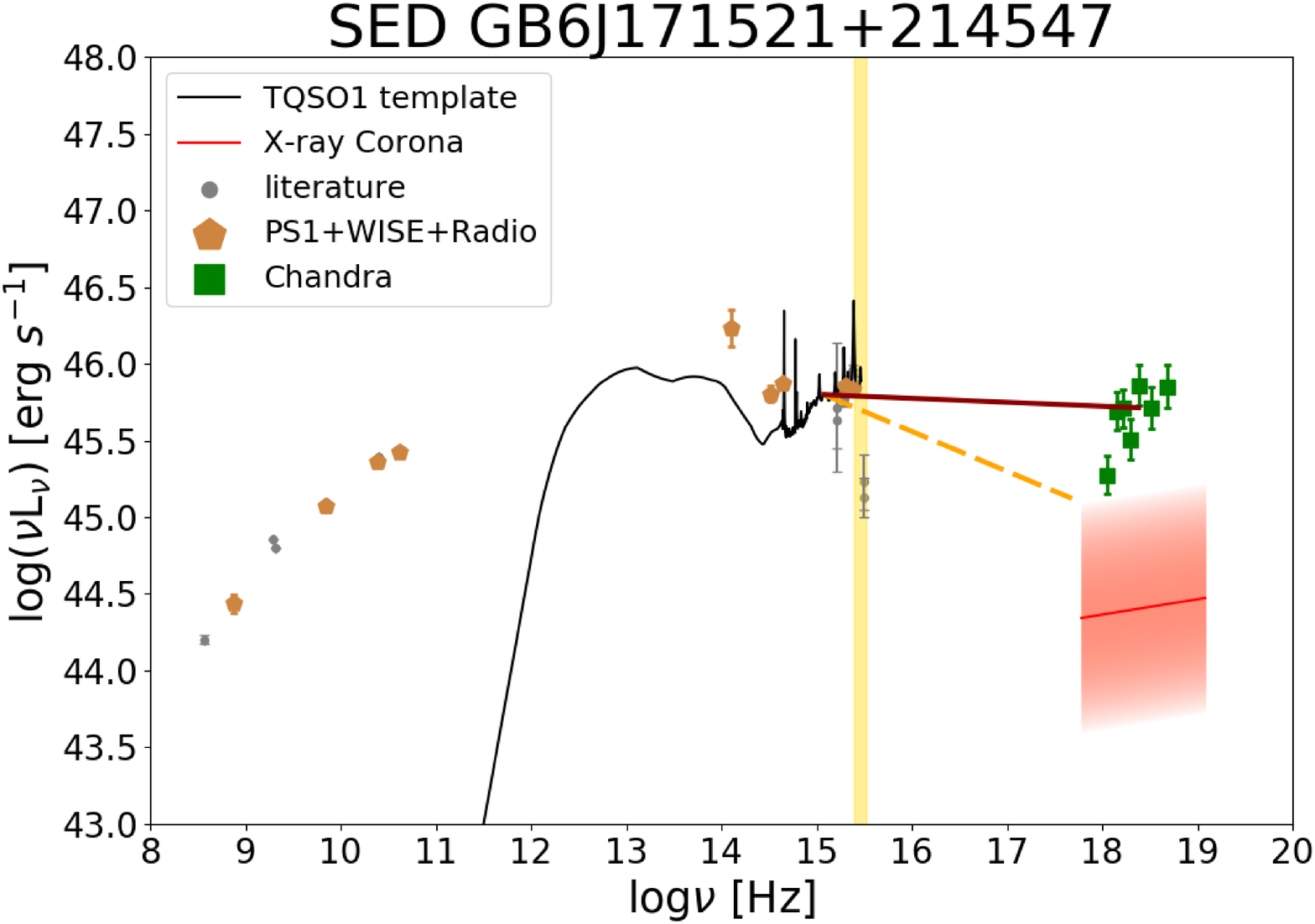}}
\sk
\subfigure{\label{2314}\includegraphics[width=0.48\linewidth]{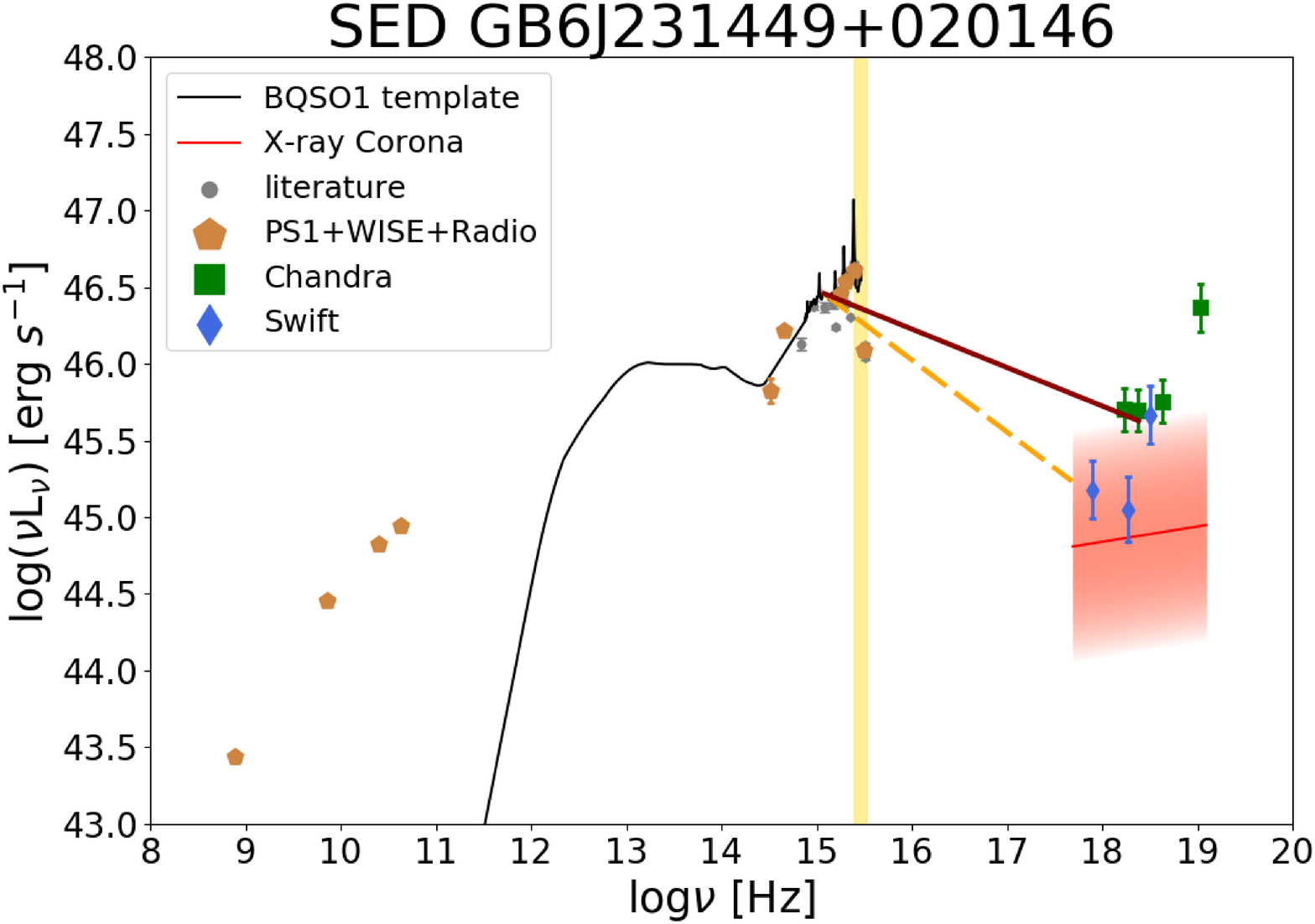}}
\subfigure{\label{2357}\includegraphics[width=0.48\linewidth]{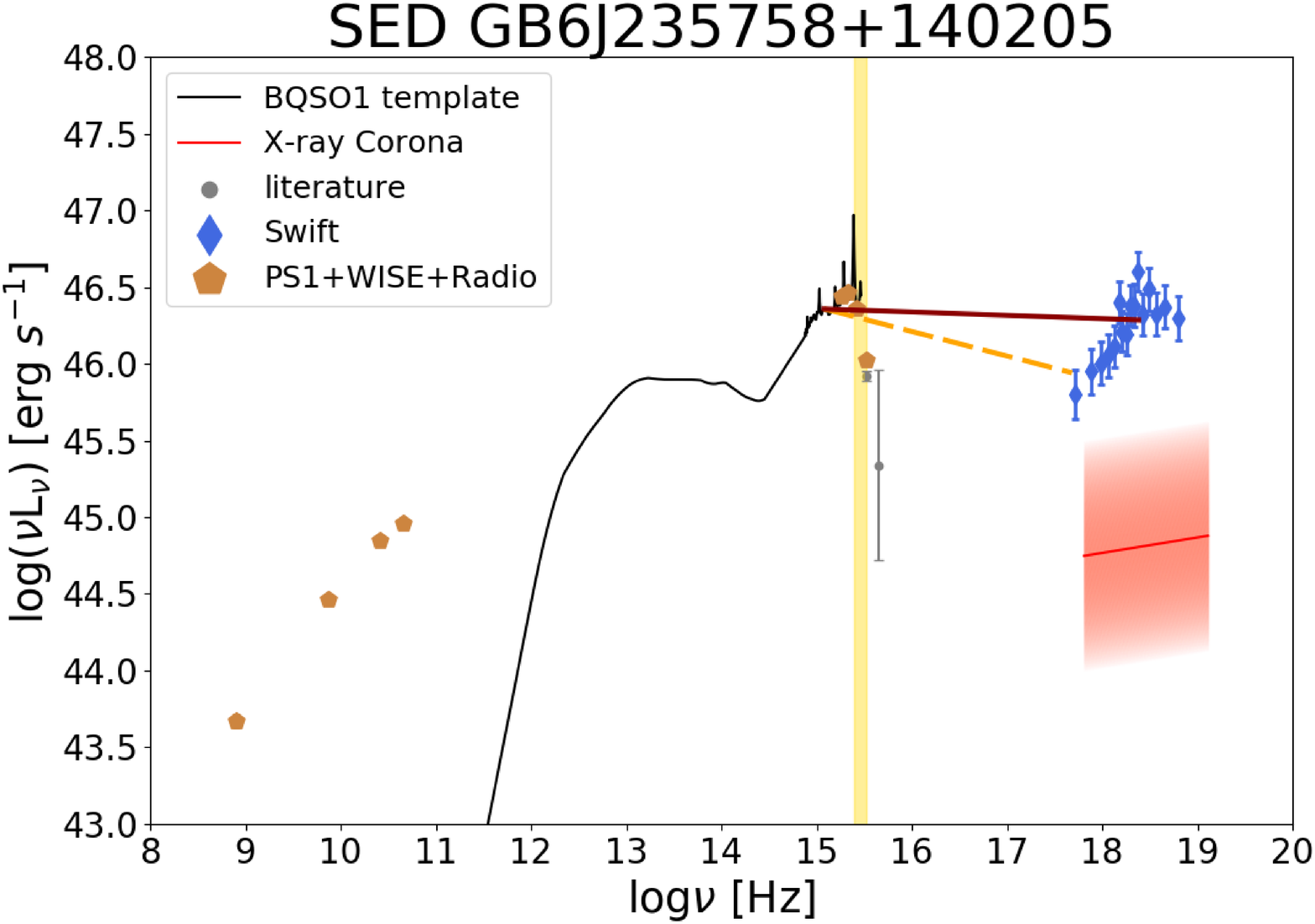}}
\sk
\caption{Broadband SEDs of the sources of the CLASS sample discussed in this paper. In all the SEDs we report the expected X-ray coronal emission from a RQ AGN with similar optical luminosity (red region) and the spectral region where hydrogen absorption is relevant (yellow region). In this representation the plotted slopes of the dashed orange and the continuous red lines are equal to  1-\aox and $1-\tilde{\alpha}_{ox}$ respectively.}
\label{seds}
\end{figure*}

In this section we report and discuss all the rest frame Spectral Energy Distributions (SEDs) of the CLASS sources. They have been built by combining multiwavelenght observations from the radio up to X-ray band. X-ray data are the ones analysed in this work, whereas other photometric points have been taken from different surveys:
\begin{itemize}
	\item \textbf{Radio}: The Giant Metrewave Radio Telescope (GMRT) all-sky survey (150 MHz, \citealt{Intema2017}), NVSS (1.4 GHz), GB6 (5 GHz) and Very Large Array (VLA, 8.4 GHz);
	\item \textbf{Infra-Red}: \textit{Wide-field Infrared Survey Explorer} ($WISE$, \citealt{Wright2010}) with the four filters $W1$, $W2$, $W3$, $W4$ (3.4, 4.6, 12, and 22 $\mu$m respectively). Five of our sources have not been detected in this survey;
	\item  \textbf{Optical}: Pan-STARRS1, with the filters $g$, $r$, $i$, $z$, $Y$, (4866-9633 \AA);
	\item  \textbf{Multiwavelength}: data points taken from the literature using the online tool SED builder\footnote{\url{http://www.asdc.asi.it/articles.php?id=11}} (grey points). They are mainly Sloan Digital Sky Survey (SDSS, \citealt{York2000}) and VLA Low-frequency Sky Survey Redux (VLSSr, \citealt{Lane2014}) observations.
\end{itemize}
We built the SED for all the objects in the sample (see Fig \ref{seds}) in order to visually show the intensity of the X-ray emission with respect to the optical/UV emission and to compare it to the emission expected for RL AGNs seen at large viewing angles (expected to be comparable to the RQ coronal emission, due to the de-beaming of the jet emission, e.g. \citealt{Galbiati2005}) with similar optical properties. Therefore, in all the SEDs we report the coronal X-ray luminosity expected in an AGN with similar optical properties following the relation found in \cite{Steffen2006} between the monochromatic luminosity at 2500 \AA \, and the one at 2 keV (assuming a photon index: $\Gamma$ = 1.9, continuous bright red line). The fading red region represents the 1$\sigma$ uncertainty associated to this estimate, while the yellow column indicates the region with a significant dropout of the luminosity caused by the Lyman  absorption (912-1216 \AA).\\
The templates, in black, are taken from the SWIRE template library\footnote{\url{http://www.iasf-milano.inaf.it/~polletta/templates/swire_templates.html}}; computed through the combination of the optical-IR spectra of an optically selected sample of type 1 QSO.  In particular, we considered the 3 QSO templates available from the library (QSO1, BSQO1 and TQSO1) characterised by different intensities of the infrared emission, and we chose the one that best represented the data\footnote{The plotted templates only serve as qualitative guide-lines to show the different components of the SED and they are not obtained through a proper fitting procedure. As described in the text, the quantitative analysis of the X-ray-to-optical luminosity ratio is done through the two-points spectral index (\aox)}. The templates have been normalised in order to match the optical data points.\\
The X-ray data are represented with three different markers, corresponding to the telescope used for the observation (Chandra = green squares, XMM = purple triangles, Swift = blue diamonds), while radio, IR and optical data are reported with brown pentagons. The two lines represent a power law connecting 2500 \AA \,  to 2 keV (orange dashed) and 10 keV (red continuous) rest-frame respectively.\\ 

\begin{figure*}

\centering
\subfigure{\label{0011}\includegraphics[width=0.48\linewidth]{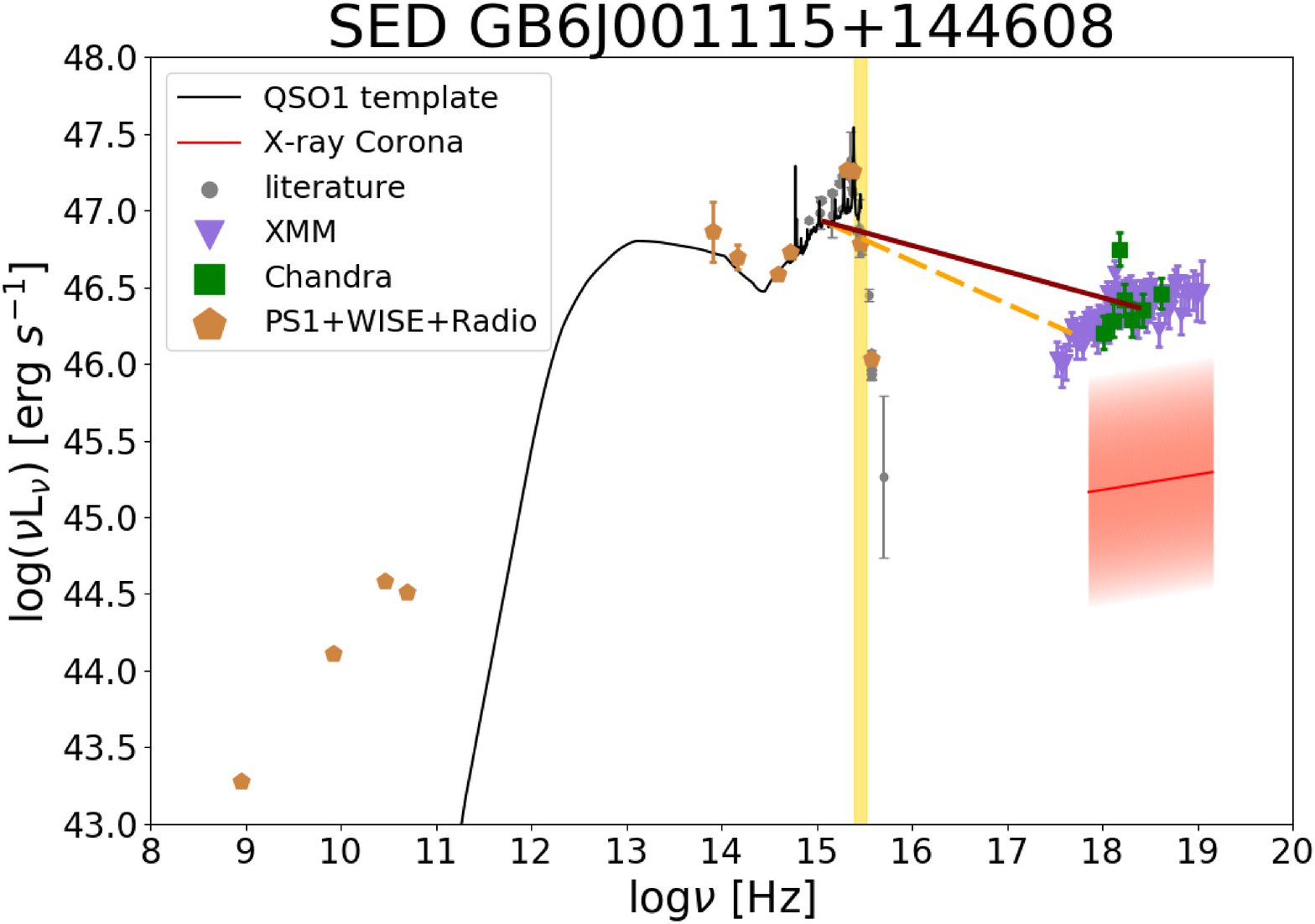}}
\subfigure{\label{0031}\includegraphics[width=0.48\linewidth]{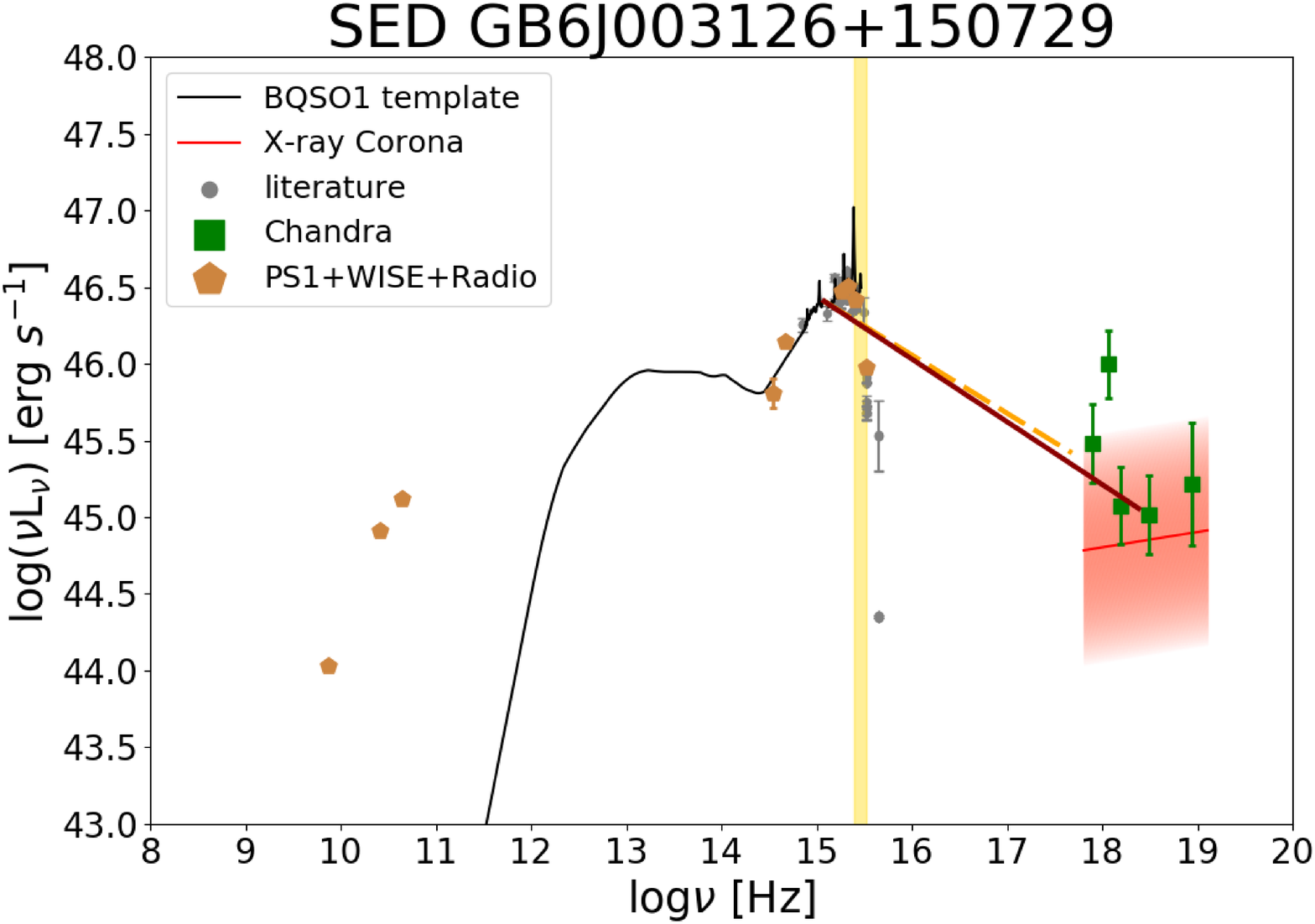}}
\sk
\subfigure{\label{0121}\includegraphics[width=0.48\linewidth]{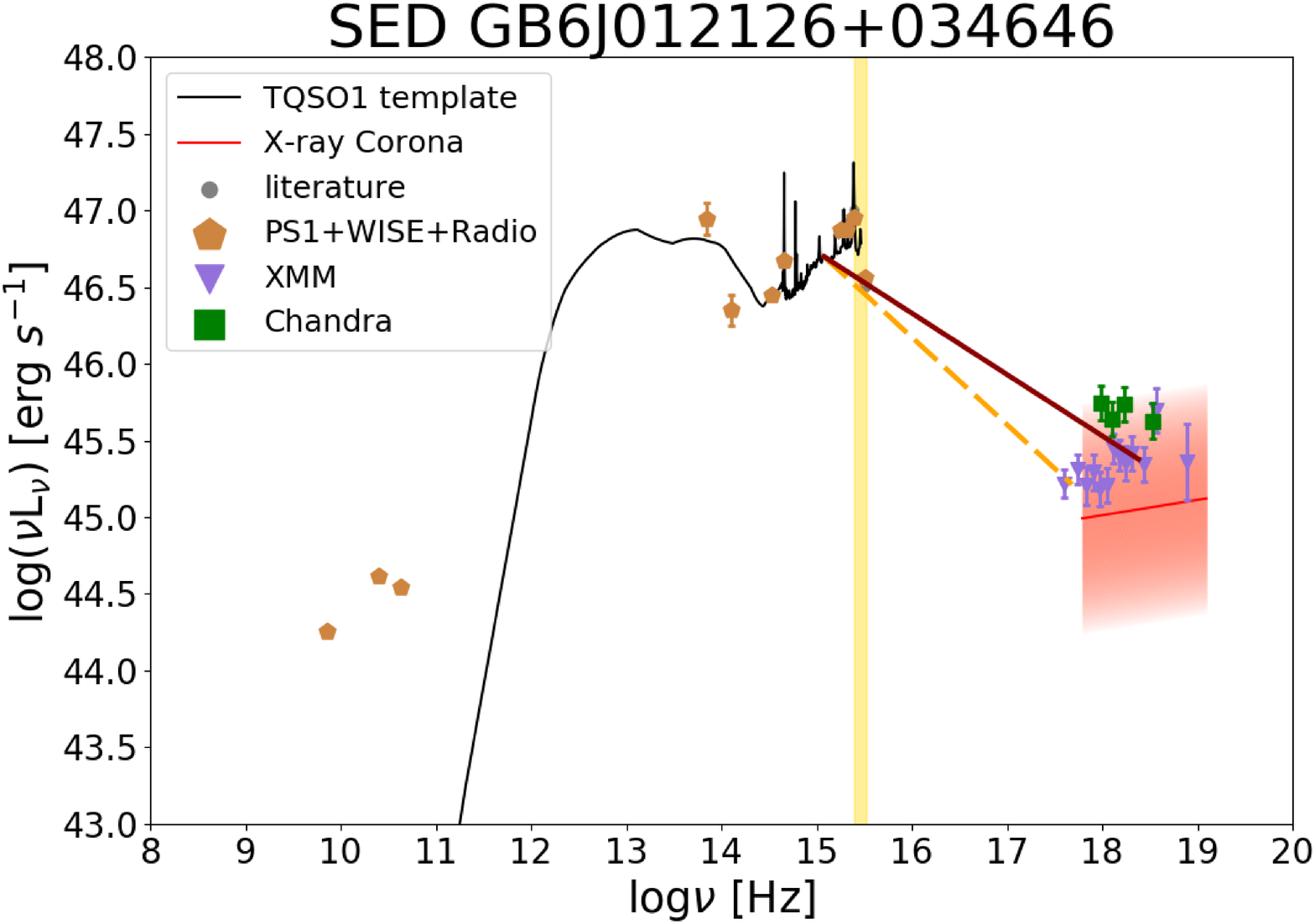}}
\subfigure{\label{0122}\includegraphics[width=0.48\linewidth]{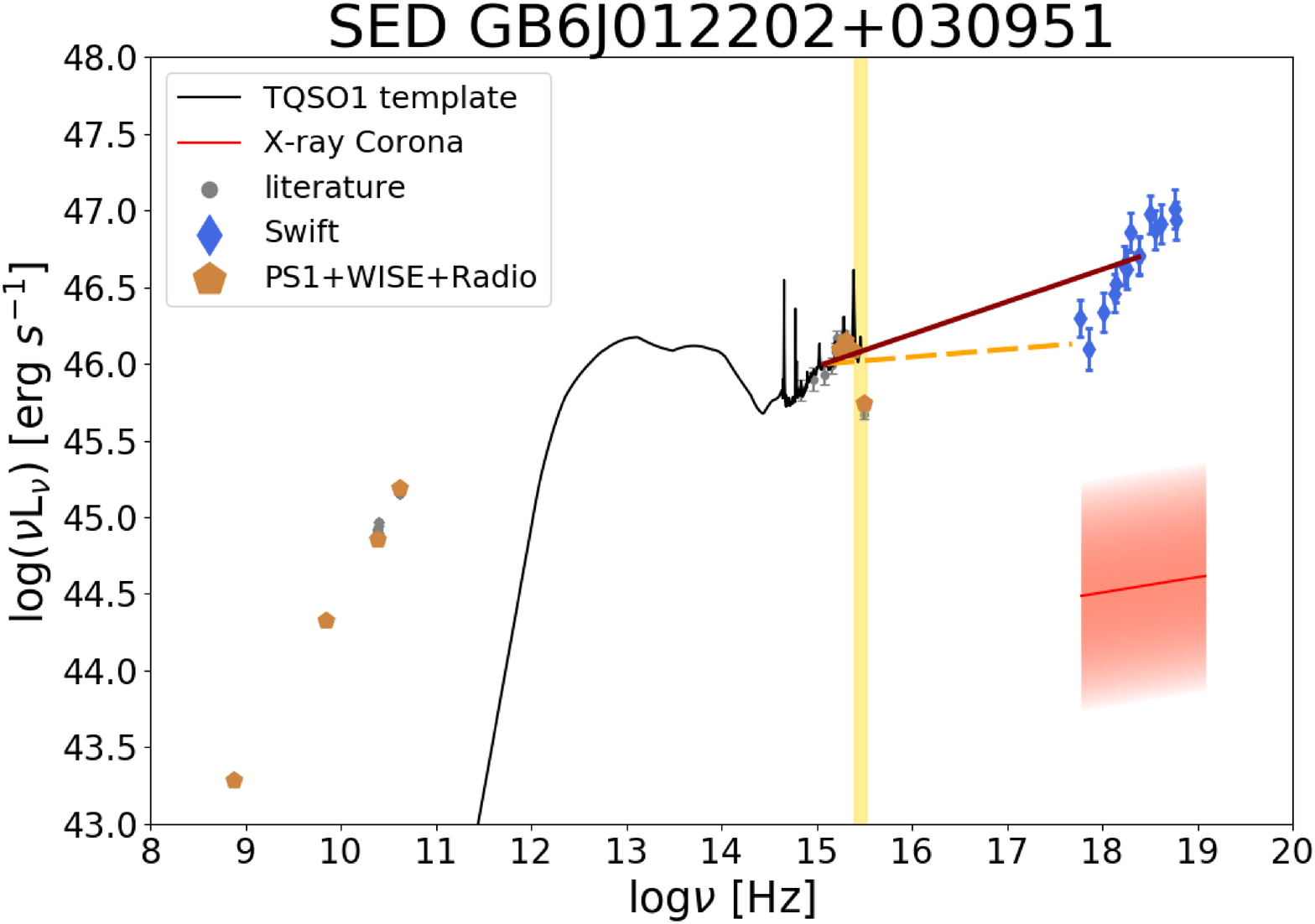}}
\sk
\subfigure{\label{0257}\includegraphics[width=0.48\linewidth]{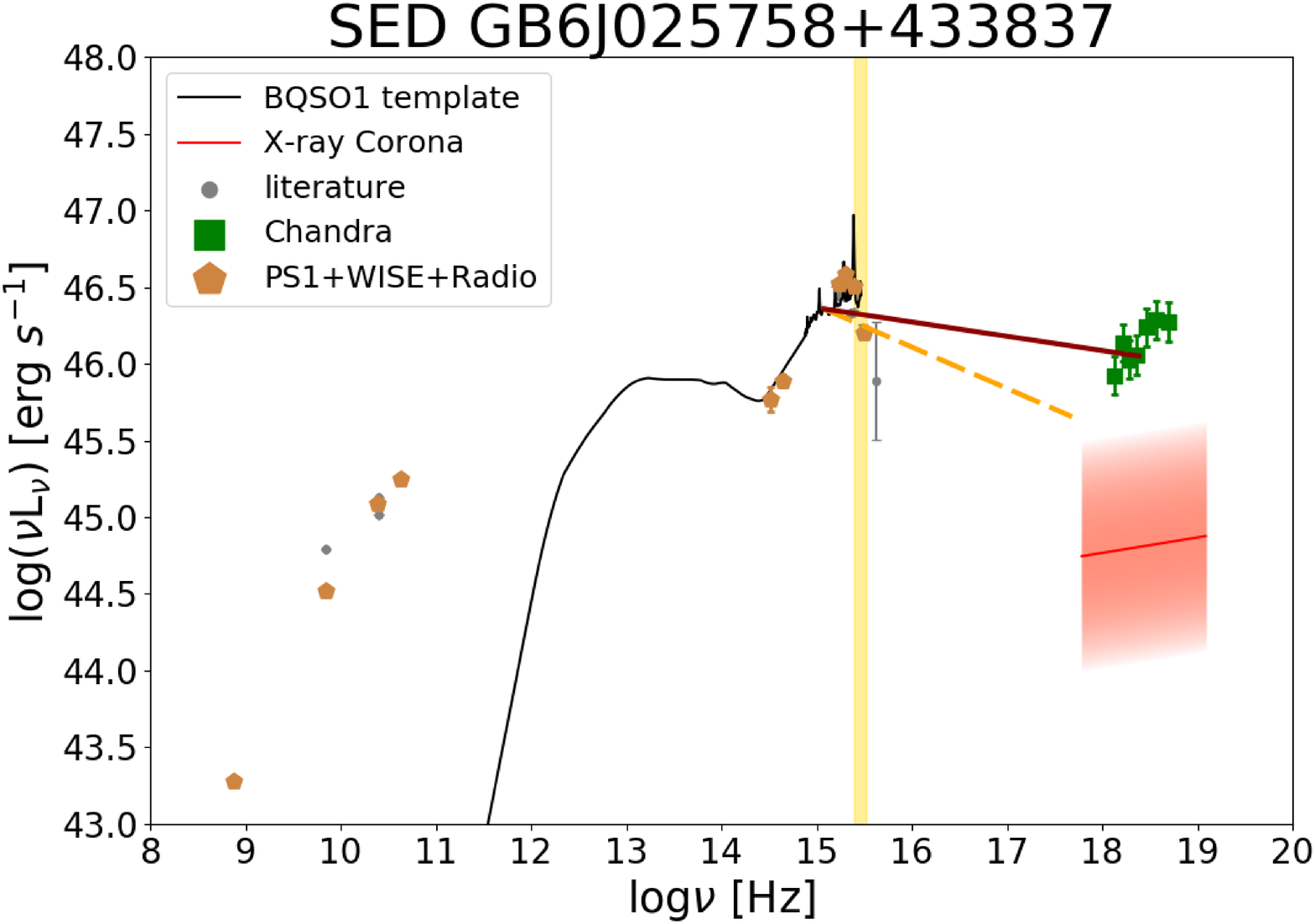}}
\subfigure{\label{0835}\includegraphics[width=0.48\linewidth]{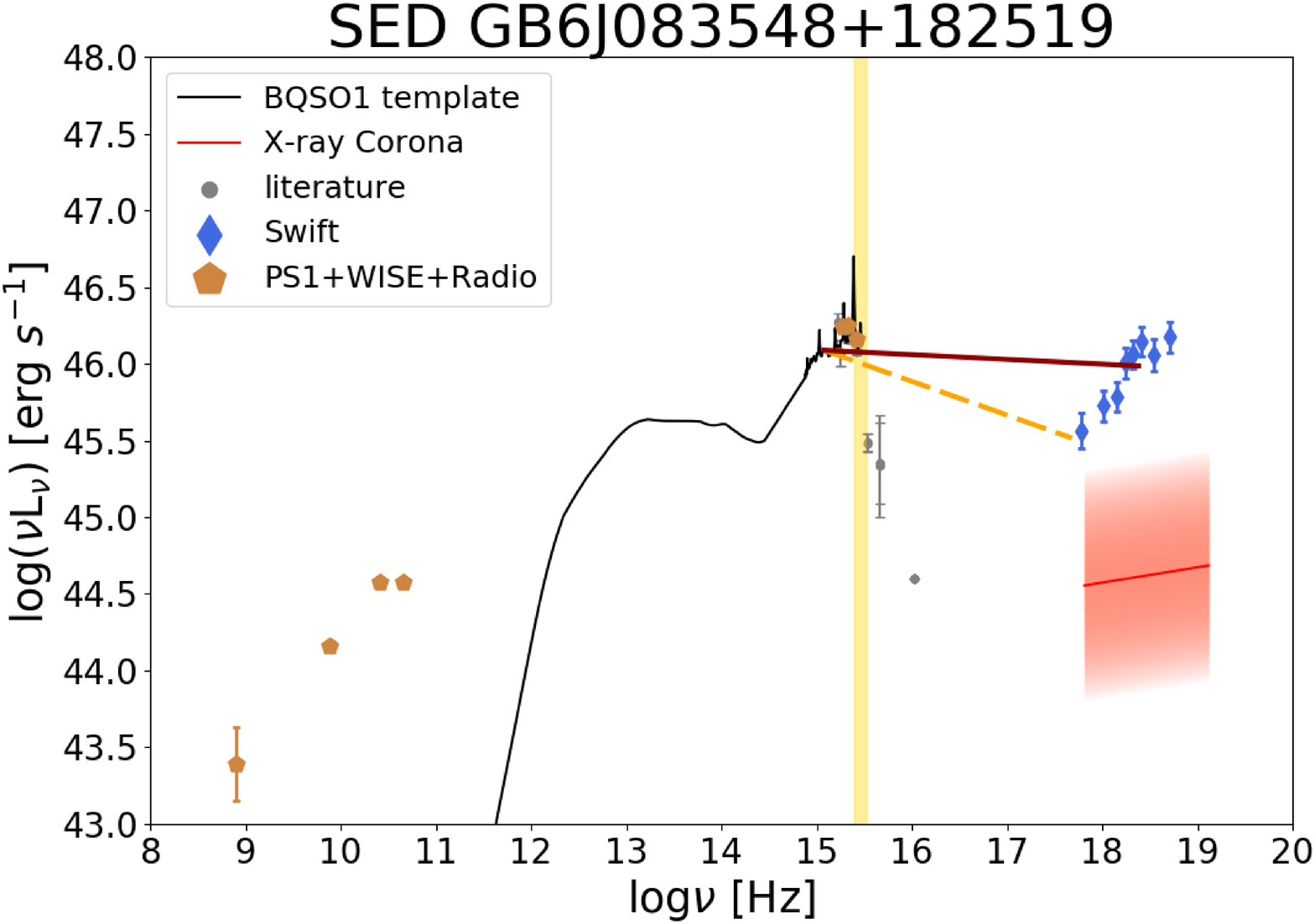}}
\sk
\subfigure{\label{0839}\includegraphics[width=0.48\linewidth]{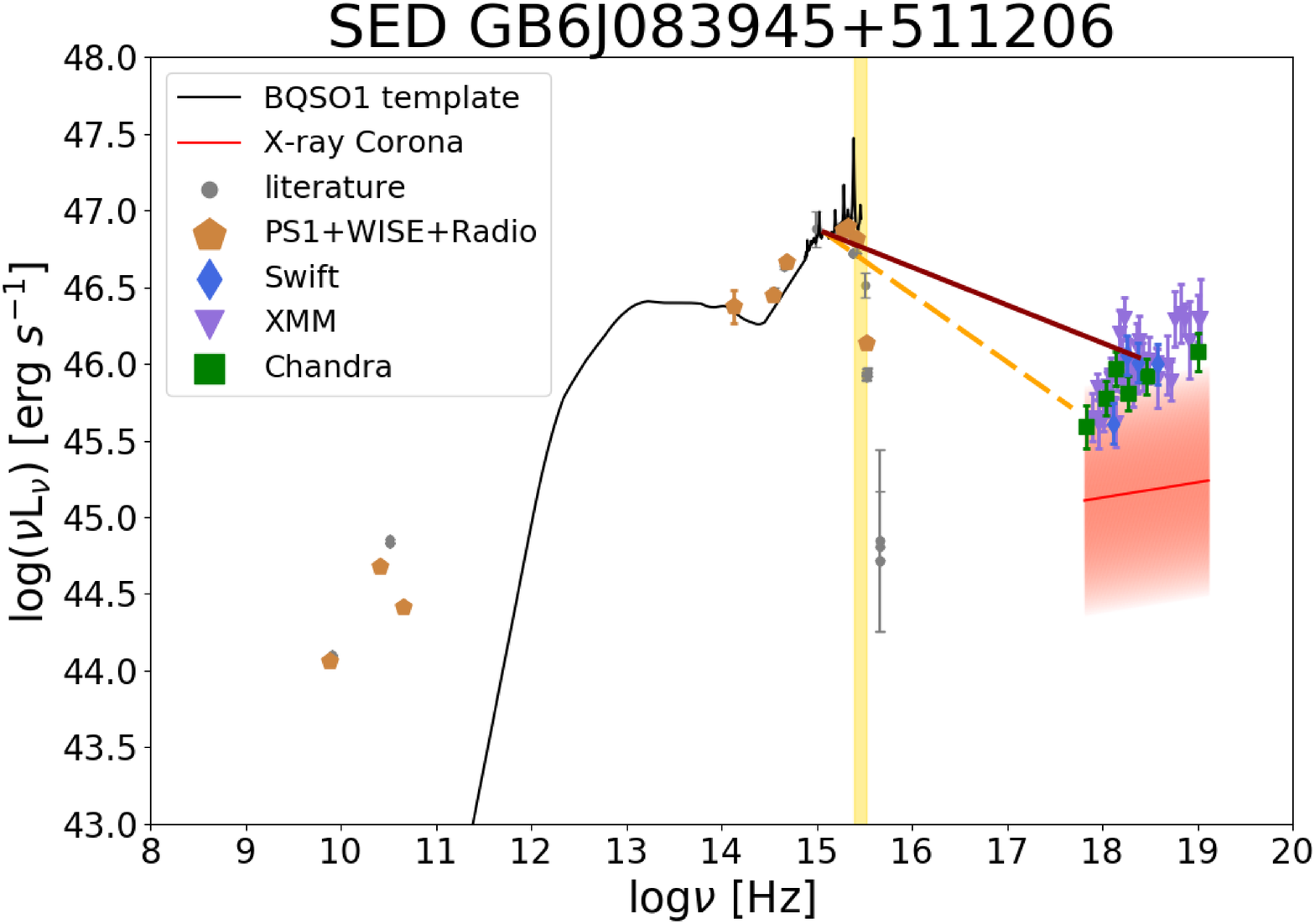}}
\subfigure{\label{0906}\includegraphics[width=0.48\linewidth]{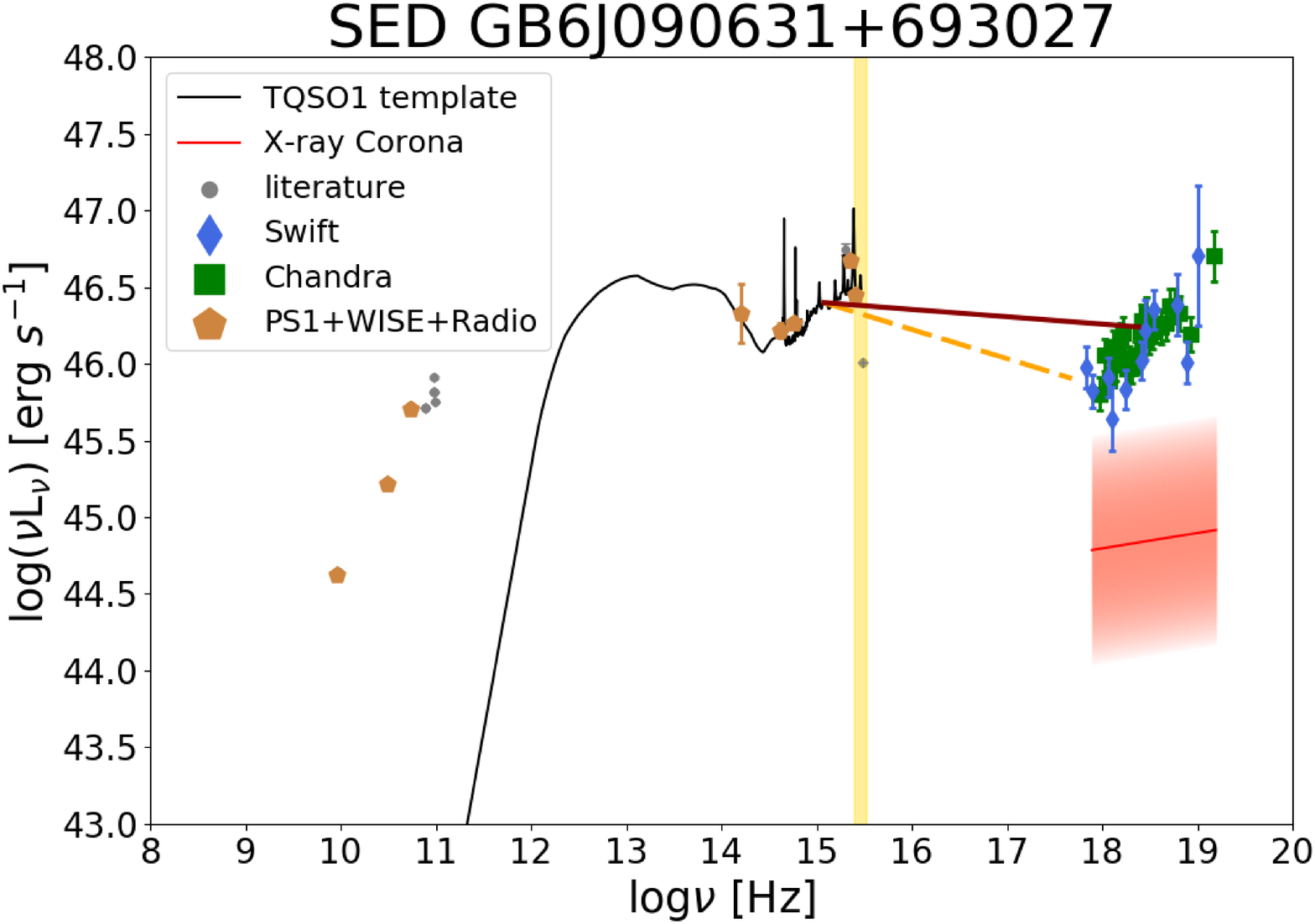}}

\end{figure*}

\begin{figure*}
\centering
\subfigure{\label{0918}\includegraphics[width=0.48\linewidth]{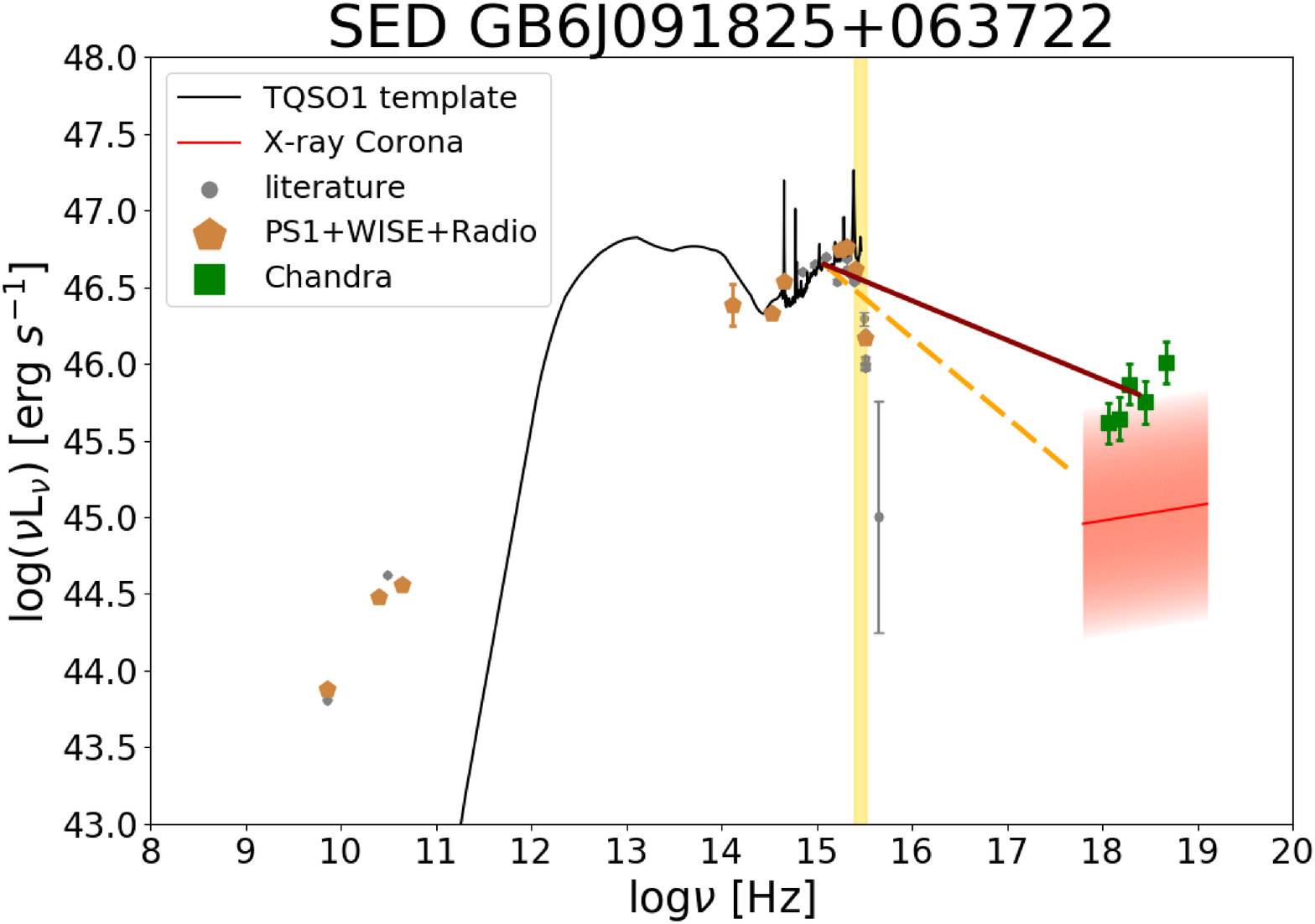}}
\subfigure{\label{1021}\includegraphics[width=0.48\linewidth]{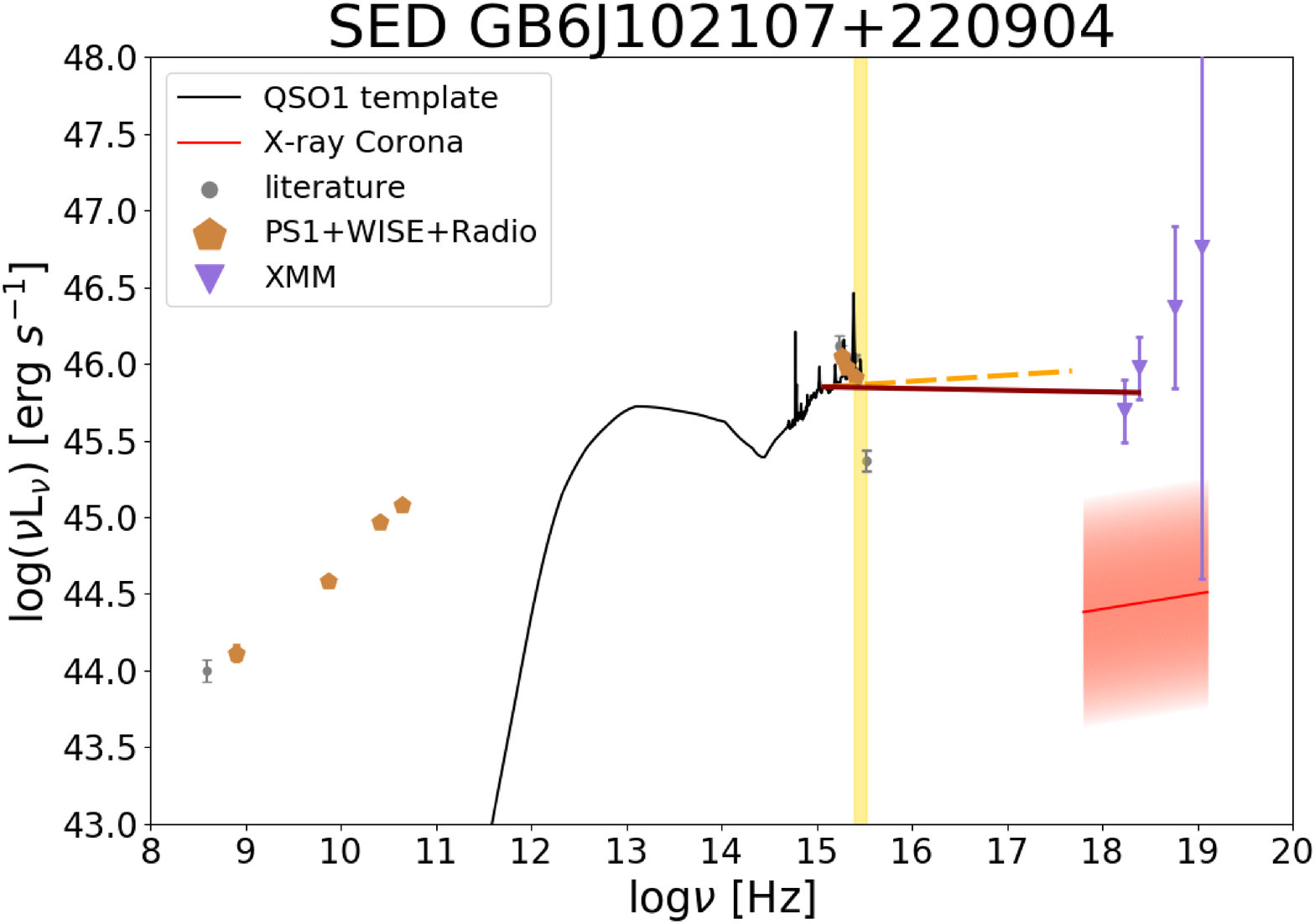}}
\sk
\subfigure{\label{1026}\includegraphics[width=0.48\linewidth]{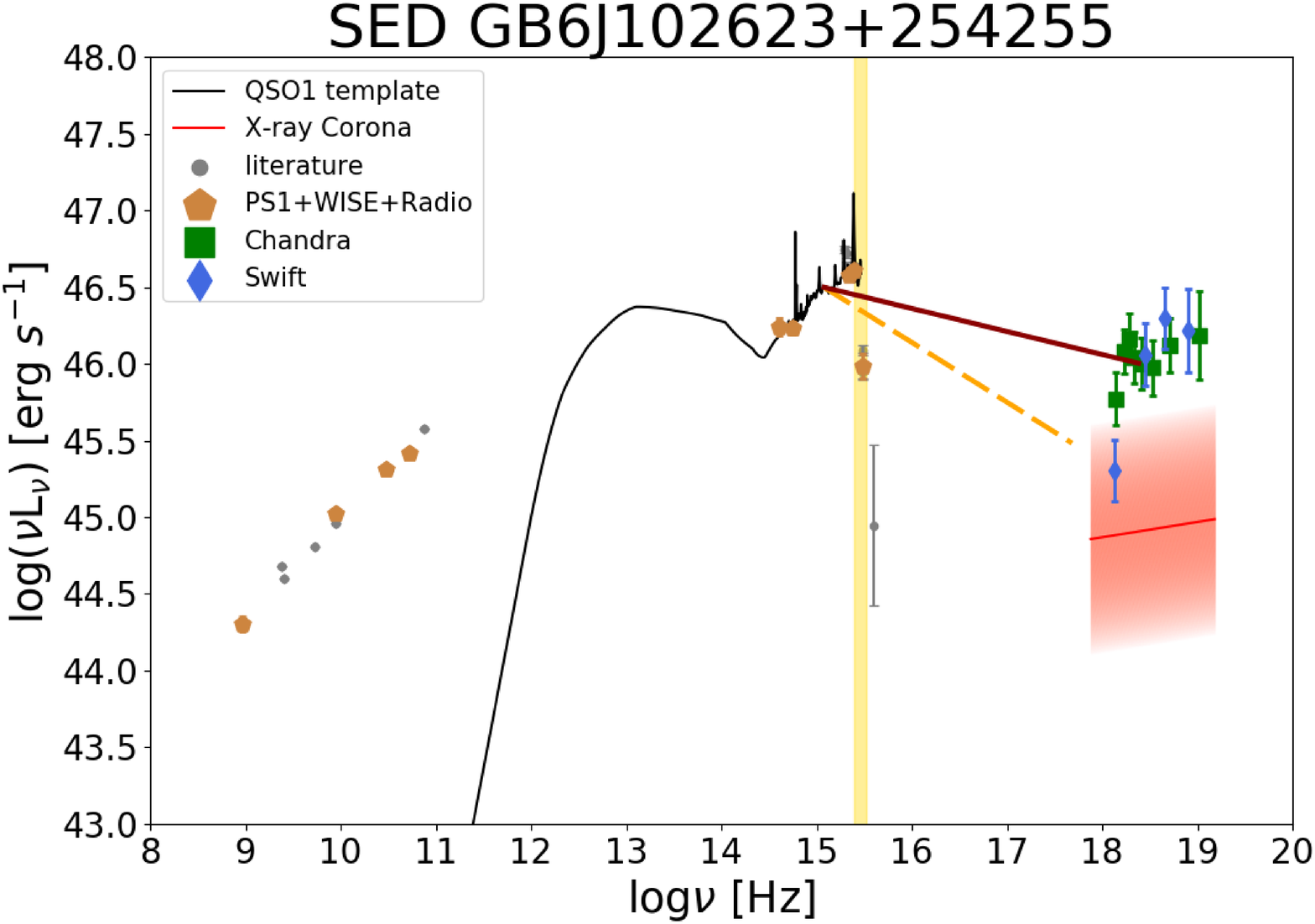}}
\subfigure{\label{1325}\includegraphics[width=0.48\linewidth]{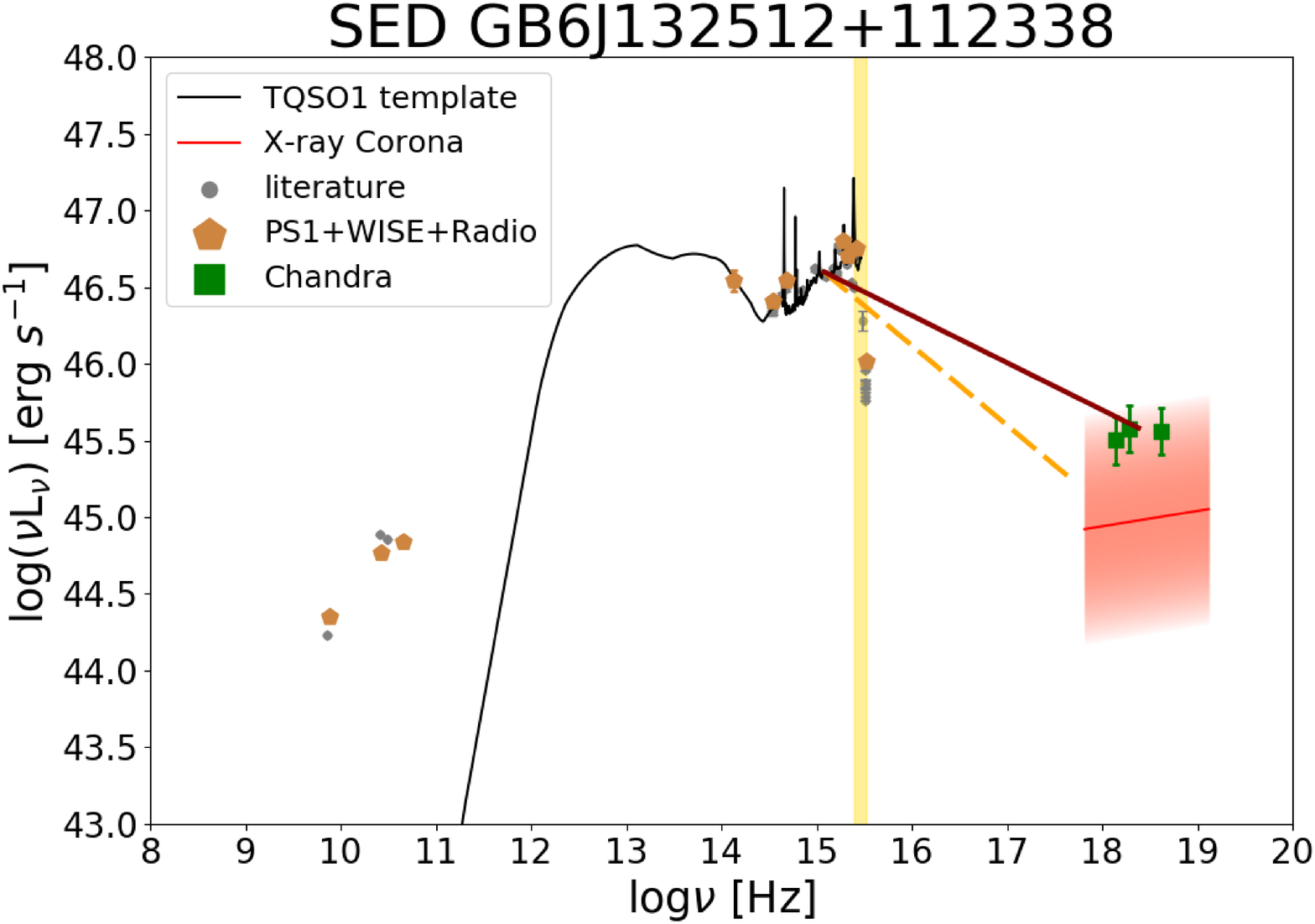}}
\sk
\subfigure{\label{1348}\includegraphics[width=0.48\linewidth]{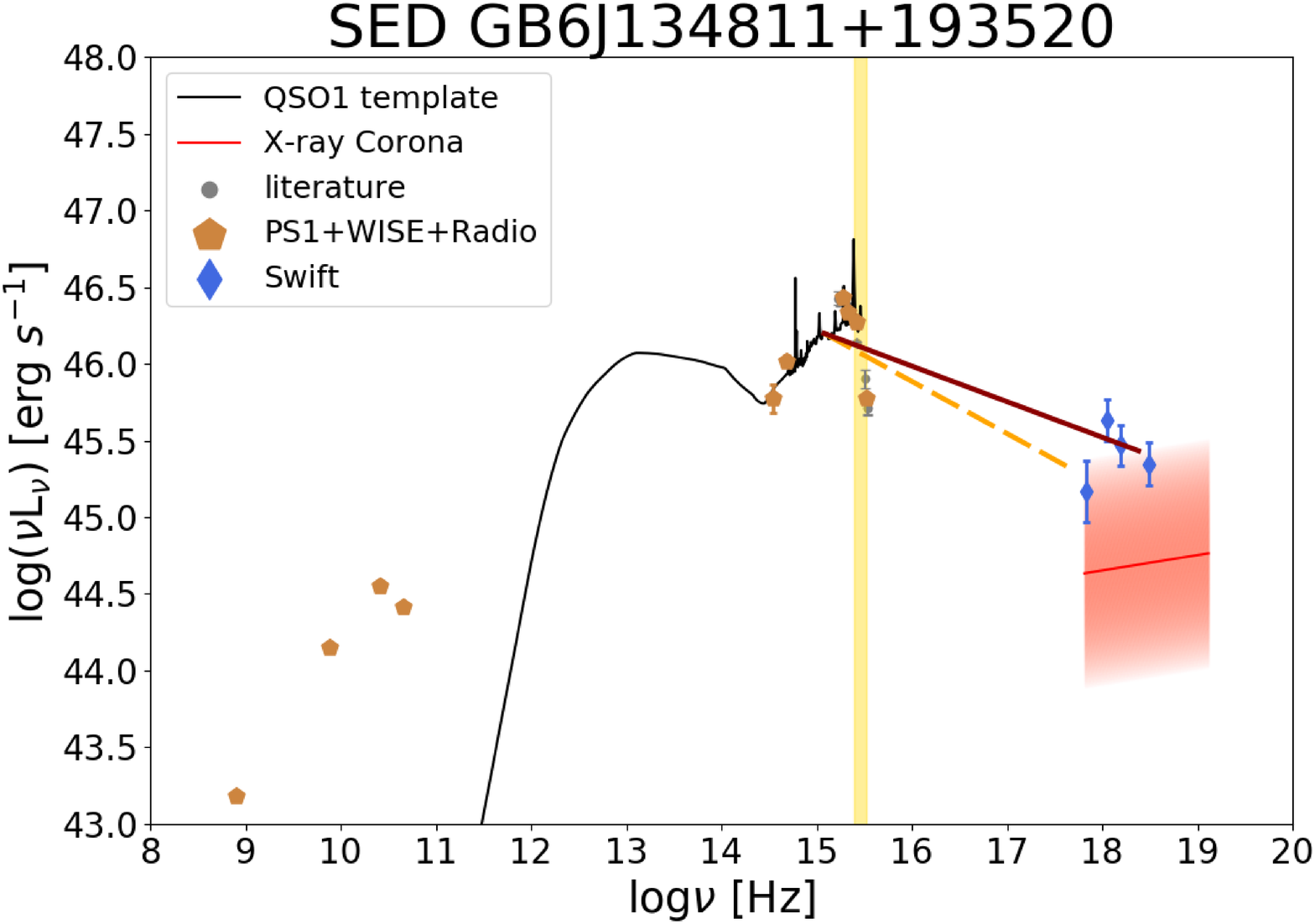}}
\subfigure{\label{1412}\includegraphics[width=0.48\linewidth]{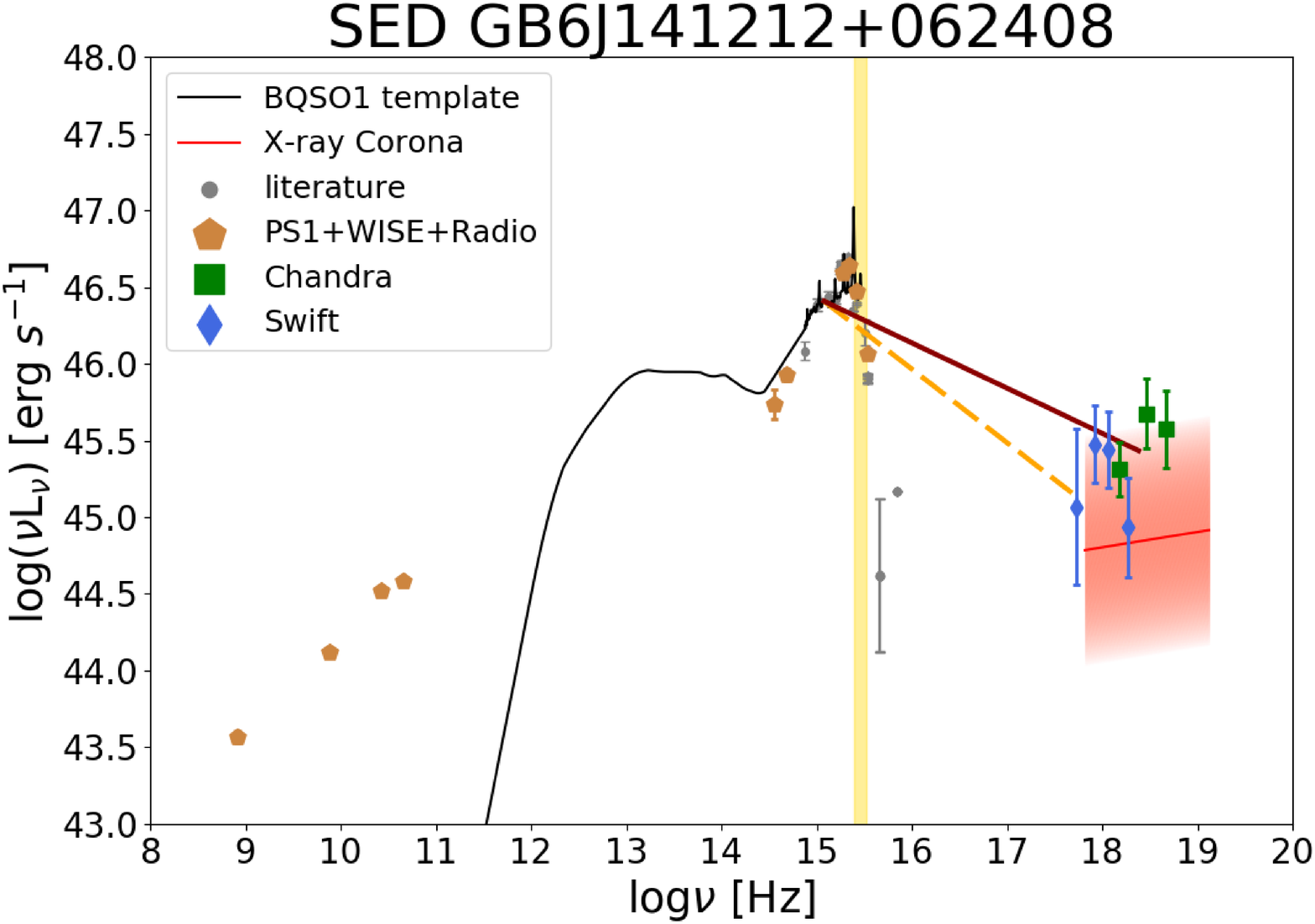}}
\sk
\subfigure{\label{1430}\includegraphics[width=0.48\linewidth]{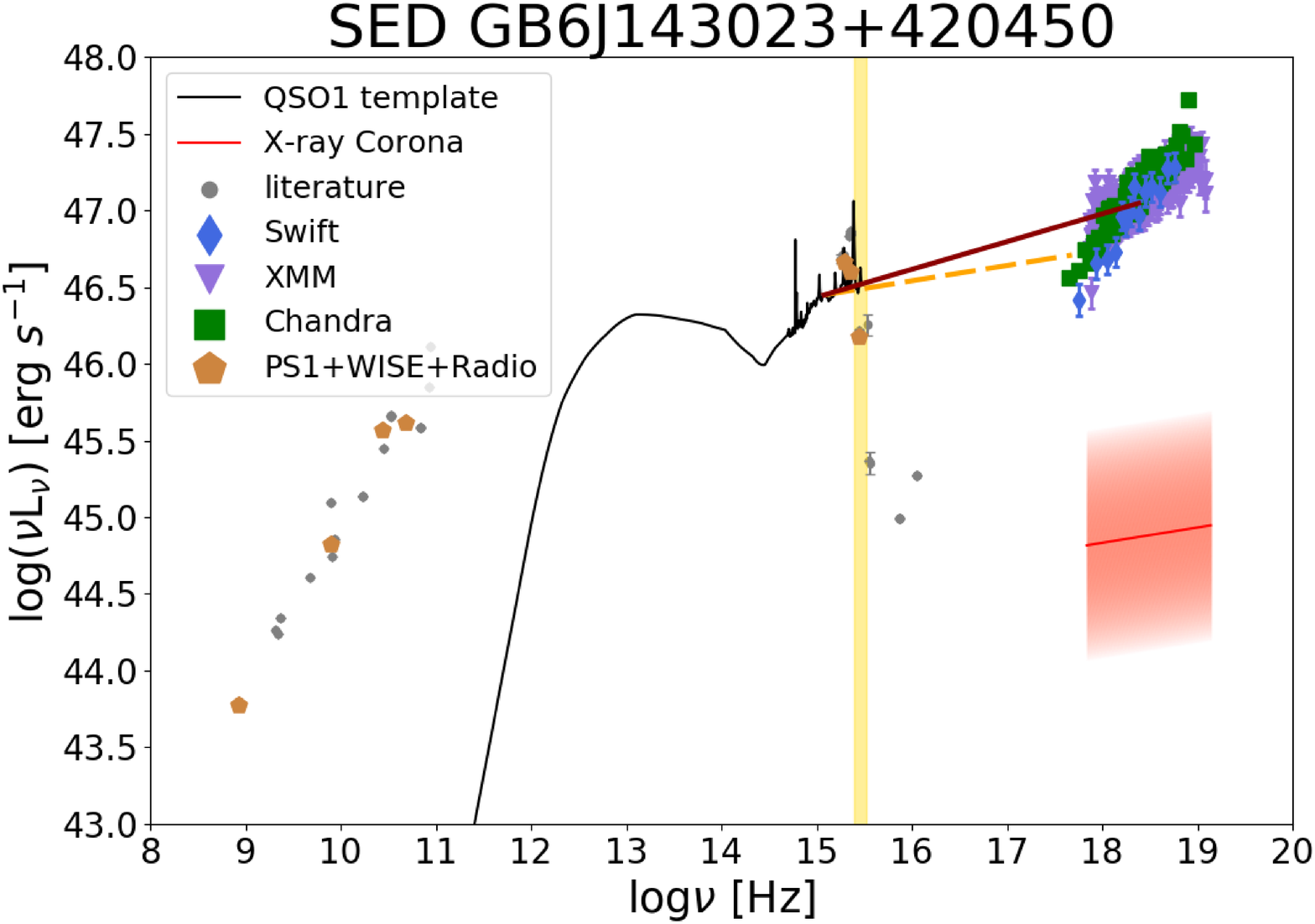}}
\subfigure{\label{1510}\includegraphics[width=0.48\linewidth]{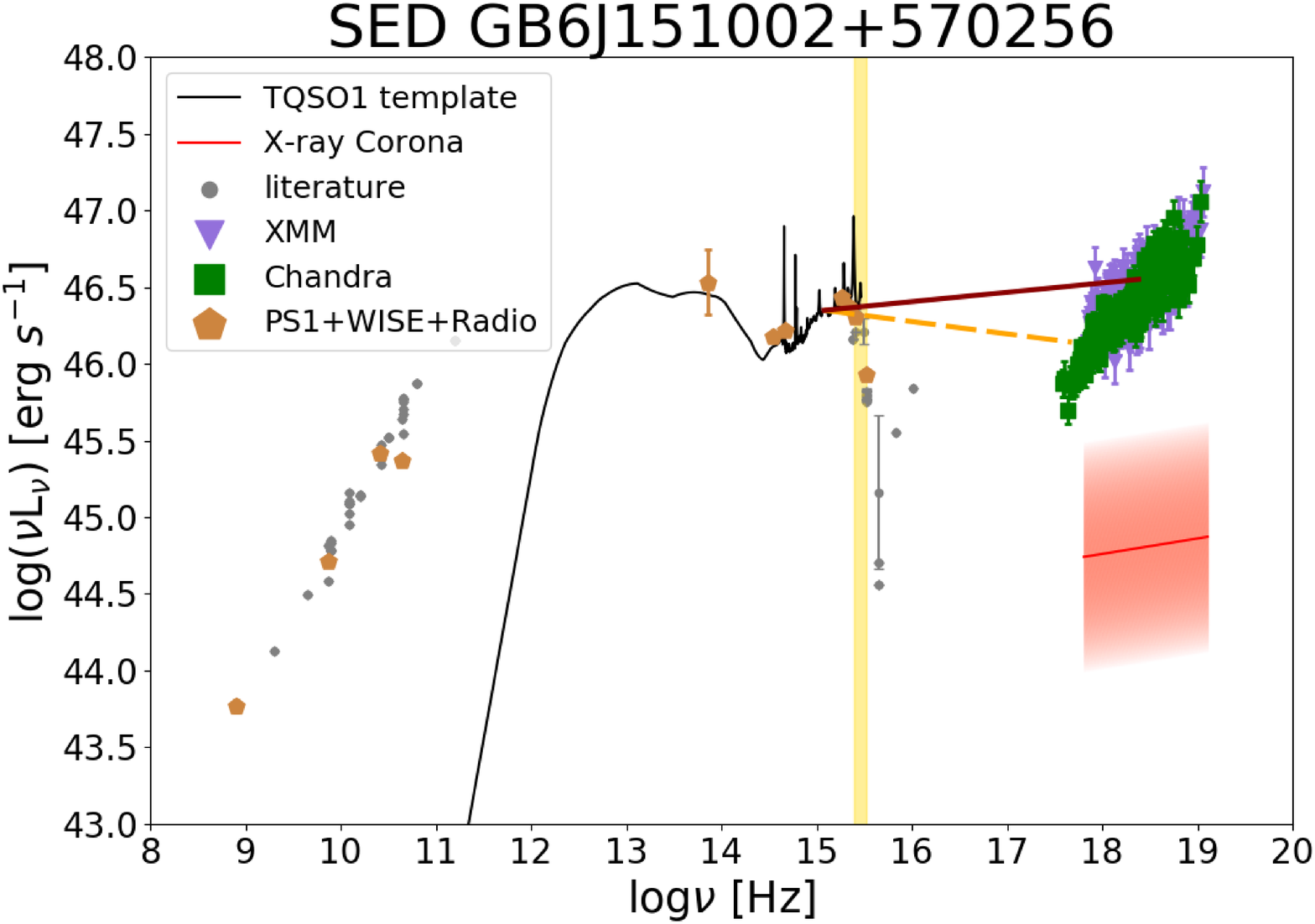}}

\end{figure*}

\section{Blazar Classification}
\label{sec:blaid}

\begin{figure*}
\centering
\includegraphics[width=0.8\linewidth]{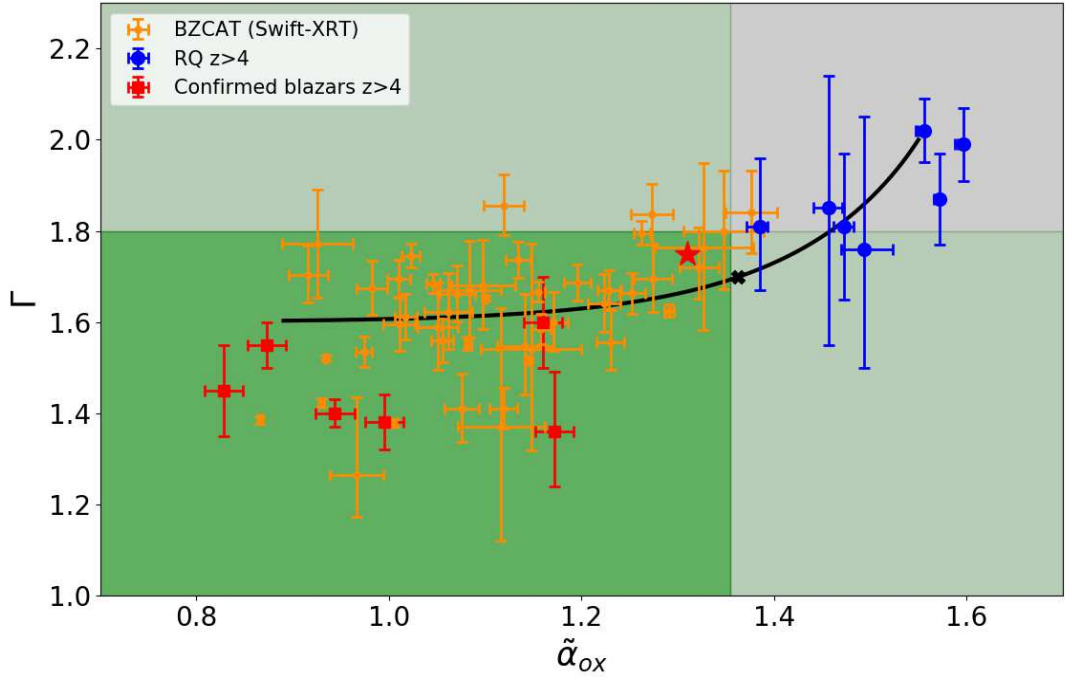}
\caption{Photon index as a function of the $\tilde{\alpha}_{ox}$ index for the comparison samples. In orange we report the BZCAT blazars observed by Swit-XRT, with red squares the few confirmed blazars at high-z and with blue points the RQ AGNs at high-z. The red star represent the z=5 blazar DESJ014132.4-542749.9, Belladitta et al. (2019). The continuous black line represents the dependence of the photon index with a variation of the X-ray intensity ($\tilde{\alpha}_{ox}$) of the jet due to beaming effects considering a coronal emission with $\tilde{\alpha}_{ox}$=1.55 and \gam=2, while for the jet emission we assume \gam=1.6. The black cross represents the critical angle assuming that the jet seen at $\theta$=0 is 50 times more intense than the corona at 10 keV. The plot is divided in four areas by a vertical line at  $\tilde{\alpha}_{ox}$ = 1.355 and a horizontal one at \gam = 1.8, which correspond to the thresholds assumed for the classification.}
\label{confirmed}
\end{figure*}

\begin{figure*}
\centering
\includegraphics[width=0.8\linewidth]{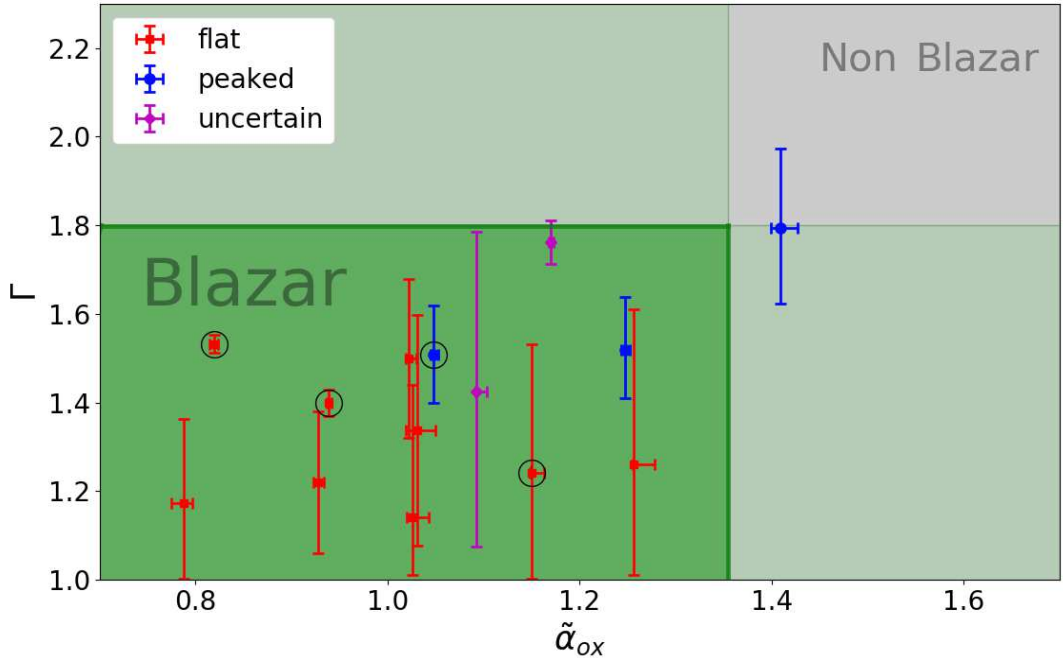}
\caption{Photon index as a function of the $\tilde{\alpha}_{ox}$  for the high-z objects in the CLASS sample with an error on the photon index lower than 0.4. The objects are plotted with different colours and markers depending on the radio spectral classification reported in C19 (``flat'' i.e. good blazar candidates, ``peaked'' i.e. probably non-blazars and ``uncertain'') and the black circles represent the few confirmed blazars at z\textgreater4 in the literature. As in the previous figure, the plot is divided in four areas by a vertical line at  $\tilde{\alpha}_{ox}$ = 1.355 and a horizontal one at \gam = 1.8.}
\label{atildeox}
\end{figure*}

The \aox \, parameter \citep{Tananbaum1979} is commonly used in the literature in order to quantify the relative strength of the X-ray emission with respect to the optical/UV component. This parameter is the two-point spectral index of a fictitious power law connecting 2500 \AA \, and 2 keV in the source rest-frame. We report the value of the \aox \, for the CLASS sample in Table \ref{tableall} and, in the SEDs  of Fig \ref{seds}, "1-\aox" \, is reported as the slope of the orange dashed line. The luminosity at 2500 \AA \, has been computed from the $i$-band magnitude (PS1) assuming an optical spectral index $\alpha_{\nu}$=0.46 \citep{Berk2001}. Both the magnitude and the luminosity are also reported in Table \ref{tableall}.\\
Blazars and mis-aligned RL AGNs can be distinguished on the basis of the value of \aox: sources with an X-ray emission strongly dominated by the relativistic jet will have a ``flat''  \aox \, index ($\lesssim$1.50, \citealt{Donato2001}), while mis-aligned objects, where most of the observed X-ray emission is due to the corona, will have a steeper (i.e. higher) \aox \, ($\sim$1.69, \citealt{Shemmer2006}). 
However, considering the monochromatic luminosity at 2 keV rest frame may not be the most convenient approach when dealing with high-z sources. Indeed, for objects with z\textgreater 4 this energy is observed at $\sim$0.3-0.4 keV, where X-ray telescopes are less sensitive, making the estimate of the \aox \, less accurate and highly dependent on the exact value of the spectral slope. For this reason we decided to consider the X-ray flux at higher energies, where the number of detected photons is larger and the normalisation is less affected by a different photon index. To this end we introduce the parameter $\tilde{\alpha}_{ox}$, analogous to the \aox \footnote{The two are related as follows: $\tilde{\alpha}_{ox}$= 0.789 \aox \, + 0.212 $\alpha$, with $\alpha$ the spectral index of the X-ray emission.}, defined at 10 keV rest frame  (in Fig \ref{seds} "1-$\tilde{\alpha}_{ox}$" is reported as the slope of the continuous red line).\\
\begin{equation}
\tilde{\alpha}_{ox} \, = - \, \frac{ log(L_{10keV} \, / \, L_{2500\angstrom})}{log(\nu_{10keV} \, / \, \nu_{2500\angstrom})} = -0.3026 \, log(\frac{L_{10keV}}{ L_{2500\angstrom}})
\end{equation}
\\
The second parameter we used for the blazar classification is the photon index. Indeed, we expect to observe different values of photon index, flat ($\lesssim$1.8) for blazars and relatively steep ($\sim$1.9) for non blazars (e.g. \citealt{Giommi2019}).
We therefore used the two parameters, $\tilde{\alpha}_{ox}$ and \gam, for the blazar classification of the CLASS sources.
In order to calibrate the classification of the sample, we decided to use two reference samples taken from the literature. As first sample, we selected all the FSRQs present in the 5$^{th}$ BZCAT edition (\citealt{Massaro2015}) with a radio flux density exceeding 1.5 Jy at 1.4 GHz. The reason for imposing the large flux limit is many-fold: first, at these flux levels, almost all of the blazars have already been  discovered and, therefore, this can be confidently considered as a radio flux limited sample (like CLASS). Second, the large majority of these objects has already been  observed (and detected) in the X-rays. This is important to avoid the introduction of possible biases against X-ray weak blazars. For these reasons, this sample should be reasonably representative of the blazar population. Finally, with this flux limit we select objects in a similar range of radio power as the CLASS sources. To have an estimate of their X-ray slope and flux, we analysed all the Swift-XRT observations that are available for the majority (60 out of 105) of these blazars using the same model adopted for the CLASS, i.e. a Galactic absorbed power law. We then considered only the objects with an optical counterpart in the PS1 catalogue (47) in order to compute the $\tilde{\alpha}_{ox}$. The photon index as a function of the $\tilde{\alpha}_{ox}$ for these objects is reported in Fig \ref{confirmed} (orange points). In this figure, we also report the few confirmed  blazars at z\textgreater 4 with accurate determination of the X-ray parameters (taken from the literature) as red squares, together with the best-fit values of the faintest high-z (z\textgreater5) blazar known so far, i.e. DESJ014132.4-542749.9 (\citealt{Belladitta2019}, red star). On the other hand, to have a term of comparison also for the coronal emission, we considered the sample of  high redshift (z\textgreater 4) RQ AGNs discussed in \cite{Shemmer2005}, since, as already mentioned, the X-ray-jet emission of RL AGNs is expected to be similar to RQ AGNs. Also in this case, we considered only the sources with an accurate estimate of the X-ray properties (photon index error\textless0.3). These sources are reported in Fig \ref{confirmed} as blue points. The continuous black line in Fig \ref{confirmed} is the predicted dependence of the X-ray slope with the  $\tilde{\alpha}_{ox}$ according to the beaming model and assuming that unbeamed objects have $\tilde{\alpha}_{ox}$=1.55 and $\Gamma$=2 while the jet has $\Gamma$=1.6. The black cross indicates the critical angle that discriminates blazars from non-blazars (1/$\Gamma$, with $\Gamma$ the bulk Lorentz factor of the jet), under the assumption that the intrinsic X-ray luminosity ratio (at 10 keV) between the blazar jet (viewing angle $\theta$=0) and the corona is 50, similar to the maximal ratios observed in the CLASS sample (this critical angle can change with a different normalisation). We have used a Lorentz factor of 10, but there is very little dependence from the assumed value. The difference between the two populations is clear: RQ AGNs occupy only the top-right region of the plot, meaning that they have a weak and steep X-ray emission. On the other hand, most blazars have a stronger and flat emission and they are located in the bottom-left region. Based on this distinction, we set two thresholds to differentiate blazars and non-blazar AGNs. In particular, as shown in Fig \ref{confirmed}, we adopt the values \gam = 1.8 and $\tilde{\alpha}_{ox}$ = 1.355 as thresholds to separate the two populations. These limits include all the confirmed high-z blazars. We then apply these criteria to the high-z AGN in the CLASS sample (Fig \ref{atildeox}). In particular, we plot here the 14 objects with a reasonable estimate of the photon index (error<0.4). The remaining 10 will be discussed further below. The sources of our sample which have  been already confirmed as blazars  in the literature are highlighted by a black circle. In addition, in the plot it is also included the classification as blazar/non-blazars (flat/peaked) based on the radio spectrum and discussed in C19 (blue points = peaked, red squares = flat and purple diamonds = uncertain).\\
Considering the objects plotted in Fig \ref{atildeox}, there is only one source whose \gam and $\tilde{\alpha}_{ox}$ are not consistent with a blazar nature (GB6J012126+034646), meaning that its X-ray spectrum is too steep and too weak to be produced by an oriented jet (grey region). On the other hand, as expected, the majority of the sources (13) are in the bottom-left region (green), meaning that they are strong and flat enough to suggest that they are bona-fide blazars. Moreover, this method consistently classifies the already confirmed blazars at high redshift, including GB6J090631+693027 (discussed earlier), in spite having a peaked radio spectrum. For the remaining 10 sources of the sample, the X-ray analysis did not provide a photon index accurate enough to be used in the previous classification. In these cases we used only the information related to the X-ray intensity compared to the optical one ($\tilde{\alpha}_{ox}$) for the classification. In particular we considered as blazars the 8 sources with an  $\tilde{\alpha}_{ox}$  below the threshold ($\tilde{\alpha}_{ox}$\textless1.355), while the candidates above the limit as non blazars (2 objects). The final results of the classification are reported in Table \ref{tableall}.\\
In summary, from this analysis a total of 21 objects are consistent with a blazar nature while the remaining 3 sources do not seem to be powered by a relativistic jet oriented towards us, corresponding respectively to $\sim$90\%  and $\sim$10\% of the entire sample. There is only one source (GB6J164856+460341) that, even after the introduction of the new parameter, keeps a relatively large uncertainty on the X-ray intensity. Given the large uncertainty on the $\tilde{\alpha}_{ox}$ value and its proximity to the adopted threshold, we classify this object as "blazar?". Also its classification as "flat radio source" is uncertain. A firm classification of this object is particularly interesting given its very high redshift (z=5.36) that would make him the second most distant blazar discovered so far. We are observing this object with VLBI to secure a firmer classification.\\
This classification has a relatively good correspondence with the one based on the radio spectra from C19. In particular, the majority ($\sim$90\%) of the candidates classified as blazars in C19 ("flat") has been confirmed also by the X-ray analysis. On the other hand, as mentioned before, there is a significant number of sources (5) that, like the blazar GB6J090631+693027, have a peaked radio spectrum, but the X-ray analysis suggests a blazar nature. We consider the X-ray classification more reliable and, for this reason, we adopted it in the analysis presented in the next sections.\\

\begin{landscape}

\begin{table} 

\setlength{\tabcolsep}{6pt}
\centering

\caption{Multiwavelenght data of the CLASS sample, together with their X-ray and radio classification.}

\begin{tabular}{llccccccccccccc}

\textbf{Name} &      & \textbf{z} &\boldmath{$N_H$}& \boldmath{\(\Gamma\)} & \boldmath{\(\alpha_{r}\)}    & \boldmath{$m_i$} &\boldmath{$log(L_{2500\angstrom}$)} & \textbf{log(R)}   & \boldmath{\(\alpha_{ox}\)}& \boldmath{\(\tilde{\alpha}_{ox}\)}& \boldmath{$log(\frac{L_x}{L_R}$)}&\footnotesize \textbf{X-ray class.}&{\footnotesize \textbf{Radio class.}} \\
(1)&&(2)&(3)&(4)&(5)&(6)&(7)&(8)&(9)&(10)&(11)&(12)&(13)\\
\hline
\hline
&&&&&&&&&\\
  
GB6J001115+144608 &   &4.96& 4.16 	 & 1.76$_{-0.05}^{+0.05}$	& 0.14	&18.28$\pm$0.01 	&32.02	&2.09 &1.28  $_{-0.01}^{+0.01}$ &1.170 $_{-0.002}^{+0.003}$	&3.05&  {Blazar}				&flat?		\\	
GB6J003126+150729 &   &4.29& 3.86 	 & 2.50$_{-0.59}$			& -0.53	&19.79$\pm$0.02 	&31.29	&2.64 &1.38  $_{-0.03}^{+0.15}$ &1.404 $_{-0.022}^{+0.003}$ 	&2.26&	non Blazar    			&flat		\\
GB6J012126+034646 &   &4.13& 3.13 	 & 1.80$_{-0.17}^{+0.18}$	& -0.88	&18.76$\pm$0.01 	&31.67	&2.68 &1.57  $_{-0.03}^{+0.03}$ &1.409 $_{-0.009}^{+0.018}$ 	&2.68&	non Blazar           	&peaked		\\
GB6J012202+030951 &   &4.00& 3.06 	 & 1.17$_{-0.19}^{+0.19}$	& -0.07	&20.48$\pm$0.03 	&30.96	&3.42 &0.95  $_{-0.04}^{+0.03}$ &0.789 $_{-0.013}^{+0.008}$ 	&3.05&	{Blazar}           		&flat 		\\
GB6J025758+433837 &   &4.07& 12.57   & 1.43$_{-0.35}^{+0.36}$	& -0.28	&19.45$\pm$0.01  	&31.38	&3.17 &1.27  $_{-0.09}^{+0.10}$ &1.093 $_{-0.001}^{+0.011}$ 	&2.53&	{Blazar}           		&flat?		\\
GB6J083548+182519 &   &4.41& 2.66	 & 1.34$_{-0.21}^{+0.21}$ 	& 0.21  &20.50$\pm$	0.04 	&31.03	&3.18 &1.22  $_{-0.05}^{+0.06}$ &1.031 $_{-0.011}^{+0.019}$	&2.42&	{Blazar}           		&flat		\\
GB6J083945+511206 &   &4.40& 3.31 	 & 1.52$_{-0.11}^{+0.12}$	& -0.47 &18.88$\pm$0.01	 	&31.68	&2.37 &1.44  $_{-0.03}^{+0.02}$ &1.247 $_{-0.004}^{+0.005}$ 	&2.99&	{Blazar?}           		&peaked		\\
GB6J090631+693027 &   &5.47& 3.74 	 & 1.51$_{-0.11}^{+0.11}$	&<-0.82	&20.54$\pm$0.02	&31.19	&3.33 &1.19  $_{-0.02}^{+0.03}$ &1.048 $_{-0.003}^{+0.005}$ 	&2.73&	{Blazar}\checkmark	&peaked		\\
GB6J091825+063722 &   &4.22& 2.99 	 & 1.26$_{-0.35}^{+0.35}$	& -0.17 &19.09$\pm$	0.01 	&31.56	&2.32 &1.52  $_{-0.07}^{+0.07}$ &1.257 $_{-0.003}^{+0.022}$ 	&1.76&	{Blazar?}          	 	&flat		\\
GB6J102107+220904 &   &4.26& 1.18 	 & 2.26$_{-1.64}^{+0.24}$	& 0.52	&21.09$\pm$0.06	&30.76	&3.89 &0.93  $_{-0.01}^{+0.35}$ &1.011 $_{-0.007}^{+0.034}$ 	&2.10&	{Blazar}          		&flat		\\
GB6J102623+254255 &   &5.28& 1.55 	 & 1.24$_{-0.29}^{+0.26}$	& 0.25	&20.06$\pm$0.02 	&31.35	&3.74 &1.39  $_{-0.06}^{+0.07}$ &1.150 $_{-0.002}^{+0.013}$ 	&1.40&	{Blazar}\checkmark 	&flat		\\
GB6J132512+112338 &   &4.42& 2.00 	 & 1.52$_{-0.50}^{+0.51}$	& -0.59	&19.36$\pm$0.02 	&31.49	&2.91 &1.52  $_{-0.10}^{+0.12}$ &1.308 $_{-0.010}^{+0.031}$ 	&2.41&	{Blazar?}         	 	&peaked		\\
GB6J134811+193520 &   &4.40& 1.91 	 & 1.83$_{-0.49}^{+0.55}$	& 0.00	&20.26$\pm$0.01 	&31.12	&3.08 &1.34  $_{-0.04}^{+0.13}$ &1.232 $_{-0.001}^{+0.086}$ 	&2.14&	{Blazar}           		&peaked		\\
GB6J141212+062408 &   &4.47& 2.10 	 & 1.62$_{-0.50}^{+0.49}$	& 0.43	&19.55$\pm$0.04	&31.42	&2.75 &1.48  $_{-0.08}^{+0.07}$ &1.295 $_{-0.048}^{+0.043}$ 	&1.81&	{Blazar?}        		&flat 		\\
GB6J143023+420450 &   &4.72& 1.18 	 & 1.53$_{-0.02}^{+0.02}$	& -0.08	&19.79$\pm$0.04	&31.37	&3.35 &0.90  $_{-0.01}^{+0.01}$ &0.819 $_{-0.004}^{+0.004}$ 	&3.10& 	{Blazar}\checkmark	&flat 		\\
GB6J151002+570256 &   &4.31& 1.57 	 & 1.40$_{-0.03}^{+0.03}$	& 0.02	&20.13$\pm$0.02 	&31.16	&3.56 &1.08  $_{-0.01}^{+0.01}$ &0.939 $_{-0.003}^{+0.003}$ 	&2.40&	{Blazar}\checkmark 	&flat 		\\
GB6J153533+025419 &   &4.39& 3.75 	 & 1.22$_{-0.16}^{+0.16}$	& 0.41	&20.17$\pm$0.04 	&31.16	&3.06 &1.12  $_{-0.04}^{+0.05}$ &0.928 $_{-0.005}^{+0.006}$ 	&2.68&	{Blazar}           		&flat 		\\
GB6J161216+470311 &   &4.36& 1.02 	 & 1.89$_{-0.69}^{+0.70}$	& 0.46	&20.18$\pm$0.02 	&31.15	&3.11 &1.51  $_{-0.12}^{+0.18}$ &1.387 $_{-0.002}^{+0.035}$ 	&1.65&	non Blazar         		&flat 		\\
GB6J162956+095959 &   &5.00& 4.71	 & 1.69$_{-0.67}^{+0.62}$	&0.49	&20.77$\pm$0.02	&31.03	&3.40 &1.20  $_{-0.13}^{+0.19}$ &1.094 $_{-0.003}^{+0.036}$ 	&2.06&	Blazar				&flat		\\
GB6J164856+460341 &   &5.36& 1.51 	 & 1.09$_{-0.09}^{+1.69}$	& -0.47	&20.31$\pm$0.02	&31.27	&2.86 &1.67  $_{-0.31}^{+0.11}$ &1.338 $_{-0.002}^{+0.116}$ 	&2.19&	Blazar?          		&flat? 		\\
GB6J171521+214547 &   &4.01& 4.73 	 & 1.14$_{-0.30}^{+0.30}$	& 0.35	&21.22$\pm$0.05   	&30.66	&4.53 &1.26  $_{-0.06}^{+0.04}$ &1.026 $_{-0.006}^{+0.017}$ 	&0.98&	{Blazar}          		&flat 		\\
GB6J195135+013442 &   &4.11& 13.30	 & 1.10$_{-0.49}^{+0.50}$	& -0.33	&20.14$\pm$0.05 	&31.12	&3.22 &1.31  $_{-0.13}^{+0.05}$ &1.045 $_{-0.006}^{+0.009}$ 	&2.55&	{Blazar}           		&flat 		\\
GB6J231449+020146 &   &4.11& 4.72 	 & 1.43$_{-0.40}^{+0.40}$	& -0.05	&19.59$\pm$0.01 	&31.34	&3.21 &1.47  $_{-0.09}^{+0.10}$ &1.251 $_{-0.006}^{+0.018}$ 	&1.83&	{Blazar}        	 	&peaked?	\\
GB6J235758+140205 &   &4.35& 3.41 	 & 1.50$_{-0.18}^{+0.18}$	& 0.18	&19.92$\pm$0.03 	&31.25	&3.28 &1.16  $_{-0.04}^{+0.05}$ &1.022 $_{-0.003}^{+0.008}$ 	&2.43&	{Blazar}           		&flat 		\\
\hline 
\hline
\end{tabular}

\begin{tablenotes}
\item \textbf{column 1:} Object name; \textbf{column 2:} Redshift; \textbf{column 3:} Galactic neutral hydrogen column density in unit of 10$^{20} cm^{-2}$; \textbf{column 4:} Photon Index between [0.5-10] keV. In case of multiple observations of the same source, we combined the different best-fit values with a weighted average;  \textbf{column 5:}  Radio spectral index between 0.15 and 1.4 GHz (observed frame), with S$_{\nu}\propto \nu^{-\alpha}$; \textbf{column 6:} Apparent magnitude in the $i$-band (7545 \AA) with its 1$\sigma$ error from Pan-STARRS1; \textbf{column 7:} Logarithm of the monochromatic luminosity at 2500 \AA \, rest frame in units of $erg$ $s^{-1} Hz^{-1}$; \textbf{column 8:} Logarithm of the radio loudness $R$ defined between 5 GHz and 4400 \AA \,  rest frame; \textbf{column 9:} Two point spectral index between 2500 \AA \, and 2 keV rest-frame; \textbf{column 10:} Two point spectral index between 2500 \AA \, and 10 keV rest-frame; \textbf{column 11:} Logarithm of the rest frame ratio between the X-ray  [0.5-10] keV and the radio  1.4 GHz luminosities ($\nu$L$_{\nu}$), both  in in units of  erg $ s^{-1}$; \textbf{column 12:} X-ray classification of the sources discussed in this work. The sign ``\checkmark" \, shows the blazars already confirmed in the literature, whereas the sign "?" refers to an uncertain classification; \textbf{column 13:} Classification based on the radio spectrum between 0.15 and 8.4GHz (observed frame) presented in C19. Also in this case the sign"?" denotes an uncertain classification.\\
\end{tablenotes}
\label{tableall}
\end{table}

\end{landscape} 

\section{X-ray luminosity Enhancement}
\label{sec:xrayradio}

\begin{figure}
\centering
\includegraphics[width=\columnwidth]{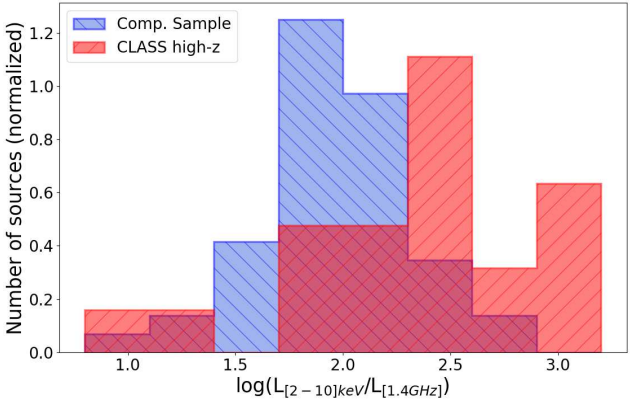}
\caption{Distribution of the ratio between the X-ray [2-10] keV and the radio 1.4 GHz  rest frame luminosities for both the CLASS sample (red) and the comparison sample of blazars at lower redshift, $\bar{z} \sim 1.1$ (blue).}
\label{istoratio}
\end{figure}

In this section we compare the X-ray-to-radio luminosity ratio (X/R ratio) of the high-z CLASS sources here classified as blazars, with the same blazar sample mentioned above (BZCAT, \citealt{Massaro2015}, S$_{1.4 GHz}$ \textgreater \, 1.5 Jy, $\bar{z}\sim 1.1$) , in order to find a possible dependence of this ratio on redshift. Figure \ref{istoratio} reports  the ratio of the integrated luminosity between [2-10] keV {(energy band directly observed in both samples)} and the radio luminosity at 1.4 GHz in the rest frame.\\
As anticipated in \cite{Ighina2018}, we find a discrepancy between the distribution of low-z and high-z blazars, in particular, the clear shift between the two peaks in Fig \ref{istoratio} suggests that either CLASS blazars have a stronger X-ray emission, or they have a fainter radio emission than low-z blazars. In order to quantify the difference in the two distributions we used the Kolmogorov-Smirnov test (KS test) on both samples. According to this test, the probability that the two samples belong to the same ``family", i.e. their distributions are drawn from a common one, is \textless 0.001\%.  In particular, the difference between the mean values of the distributions indicates that the CLASS blazars have X/R ratios $\sim$2.4$\pm0.5$ times higher than low-z blazars. Moreover we want to stress that this discrepancy is not a selection effect related to the limit imposed on the X-ray-to-optical luminosity ratio in our classification. Indeed this difference in the two distributions would still be significant (\textless 0.001\%) even considering all the sources present in the CLASS regardless to our classification as blazar and non-blazars.\\
A similar trend has been observed also by \cite{Wu2013} and \cite{Zhu2018}, in a population of very radio-loud AGN (log$R$>2.5, some in common with this work). In both works it was found a significant difference (a factor 1.9$^{+0.5}_{-0.4}$ in \citealt{Zhu2018}) in the X-ray-to-optical luminosity ratio. \cite{Wu2013} proposed as possible explanation for this trend a fractional IC/CMB model for the X-ray emission of blazars. The photons from the CMB are expected to interact via Inverse Compton (IC) scattering with the relativistic electrons in the jet producing emission in the X-rays (e.g. \citealt{Harris1979}), thus enhancing the total observed X-ray luminosity. This interaction is negligible in the compact inner-most regions of the jet (e.g. \citealt{Ghisellini2009}), but it is expected to be important for the electrons located in extended part of the jet (at few kpc), becoming negligible again at greater distances due to the deceleration of the jet (e.g. \citealt{Marshall2018}). According to this model, only a fraction of the X-ray emission observed at low redshifts is due to this interaction, the remaining being produced by IC scattering with  photons produced by the AGN itself (e.g. accretion disk, broad-line region, dusty torus, etc...). Since the density of the CMB increases as $(1+z)^4$ the interaction with these photons is expected to become more and more important at high redshifts (e.g. \citealt{Schwartz2002}),  thus producing an X-ray luminosity comparable (or even stronger) to the emission coming from the most compact region. This could explain why high-z blazars have, on average, X-ray luminosities larger than low-z blazars. In particular, following \cite{Wu2013}, we expect that:
\begin{equation}
\frac{L_X}{L_R}(z)\: = \: \frac{L_X}{L_R}(z=0)\: [1 + A (1 + z)^4]
\end{equation}
where $A$ is the fraction between the extended and the compact emission at z = 0. In order to obtain an enhancement similar to the one observed in our sample we need $A\approx$1.6$\times10^{-3}$, corresponding to a contribution of about 4\% at z = 1.3, similar to the values found in \cite{Wu2013} and \cite{Zhu2018}. In order to firmly test this interpretation, the observation and the study of the blazars in different ranges of redshift (especially at z\textgreater5.5) are necessary. It is worth noting that this effect, if present, may alter the classification of some sources made in section \ref{sec:blaid}, in particular those with a value of $\tilde{\alpha}_{ox}$ close to the threshold, leading to the mis-classification of some RL AGNs as blazars due to their enhancement of the X-ray emission. After correcting the values of the $\tilde{\alpha}_{ox}$ for the additional X-ray emission related to the CMB, there are 5 sources here classified as blazar that would overcome the threshold. For this reason, we mark the X-ray classification of these objects in Table \ref{tableall} with a "?". In any case, as already explained above, a change of classification of these sources would have a negligible impact on the analysis presented here.\\
Using the sources classified as blazar according to their radio spectrum, C19 were able to infer the space density of blazars at z\textgreater4. Even adopting the new blazar classification, based on the X-ray data, the space density discussed in C19 does not change significantly. In particular, C19 found that the space density of blazars at z>4,   including the most luminous ones, is consistent with a density evolution peaking at z$\sim$2, as suggested by \cite{Mao2017}. This is in apparent contrast with the results presented by \cite{Ajello2009} based on a sample of X-ray selected blazars (Swift-BAT), according to which the most (X-ray) luminous objects present a peak at much higher redshifts (z$\sim$4). In principle, this discrepancy could be explained by an evolution of the X-ray-to-radio luminosity ratio with redshift like the one discussed here. However, the X-ray-to-radio luminosity ratios observed in the highest redshift blazars of the \cite{Ajello2009} sample (2\textless z\textless3) are even larger than the one observed in the CLASS z>4 blazars, something that does not seem to be consistent with the CMB model, according to which the X-ray-to-radio flux ratio should monotonically increase with redshift. Either the observed dependence of the X-ray-to-radio luminosity ratio with redshift is not due to the interaction with the CMB or its impact is not the same in all the sources.  For instance, if the value of the parameter A is not unique, but it follows (as reasonable) a distribution of values, we expect that the X-ray selected blazars at redshift 2-3 constitute the (small) tail of the population with a significantly larger value of A ($\sim$0.01-0.1), where the enhancing effect of the CMB has a major role. The presence of these few, but extreme sources, would affect the evolution estimate based on the BAT survey, suggesting a density peak at very large redshift ($\sim$3-4).  We are running detailed simulations to test the validity of this scenario.\\

\section{Conclusions}
\label{sec:sum}

In this work we presented the X-ray properties of the CLASS sample of blazar candidates discussed in C19. In order to have a reliable classification of all blazars, we performed an X-ray analysis of their Chandra, XMM-pn and Swift-XRT observations. In order to classify these sources according to the intensity and the flatness of their X-ray spectra, we re-defined the commonly used \aox \, parameter, using a higher X-ray energy, 10 keV instead of 2 keV, that was more suited for low-z sources. We then used a sample of confirmed blazars at lower redshift and one of RQ AGNs at redshift similar to the CLASS sample to estimate the limits on the photon index and the $\tilde{\alpha}_{ox}$ for the classification. We concluded that 21 sources of the CLASS sample have an X-ray emission  consistent with a blazar nature, whereas the remaining 3 are too faint to be blazars.\\
Finally, we compared the CLASS high-z blazars with a sample of blazars selected at lower redshift (z$\sim$1). In particular, we found a dependence in the class of blazars of the X/R emission ratio on the redshift. Following \cite{Wu2013} we interpret this difference as due to the interaction of the electrons in relatively extended (a few kpc) regions of the jet with CMB photons through IC. A high angular resolution radio campaign is under way in order to strengthen the blazar classification of the sources and to study of the inner part of the radio jet.

\section*{Acknowledgements}
\footnotesize
We thank M. Dotti, G. Ghisellini, P. Severgnini and C. Cicone for helpful discussions. This publication has made use of data and/or software provided by the High Energy Astrophysics Science Archive Research Center (HEASARC), which is a service of the Astrophysics Science Division at NASA/GSFC and the High Energy Astrophysics Division of the Smithsonian Astrophysical Observatory. This work has used data from observations with the Neil Gehrels Swift Observatory. The scientific results reported in this article are based in part data obtained from the Chandra Data Archive. This research has made use of software provided by the Chandra X-ray Center (CXC) in the application package CIAO. This publication is partially based on results from the enhanced XMM-Newton spectral-fit database, an ESA PRODEX funded project, based in turn on observations obtained with XMM-Newton, an ESA science mission with instruments and contributions directly funded by ESA Member States and NASA.  This work has made used of data from the Pan-STARRS1 Surveys (PS1) and the PS1 public science archive, which have been made possible through contributions by the Institute for Astronomy, the University of Hawaii, the Pan-STARRS Project Office, the Max-Planck Society and its participating institutes, the Max Planck Institute for Astronomy, Heidelberg and the Max Planck Institute for Extraterrestrial Physics, Garching, The Johns Hopkins University, Durham University, the University of Edinburgh, the Queen's University Belfast, the Harvard-Smithsonian Center for Astrophysics, the Las Cumbres Observatory Global Telescope Network Incorporated, the National Central University of Taiwan, the Space Telescope Science Institute, the National Aeronautics and Space Administration under Grant No. NNX08AR22G issued through the Planetary Science Division of the NASA Science Mission Directorate, the National Science Foundation Grant No. AST-1238877, the University of Maryland, Eotvos Lorand University (ELTE), the Los Alamos National Laboratory, and the Gordon and Betty Moore Foundation. This research made use of Astropy, a community-developed core Python package for Astronomy (Astropy collaboration 2013, http://www.astropy.org). AC, AM, SB  acknowledge support from ASI under contracts ASI-INAF n. I/037/12/0 and n.2017-14-H.0 and from INAF under PRIN SKA/CTA FORECaST.




\bibliographystyle{mnras}
\bibliography{highzblazars}




\bsp	
\label{lastpage}
\end{document}